\documentclass[twocolumn,showpacs,aps,prc,superscriptaddress,fixfloat]{revtex4}
\usepackage[dvips]{graphicx}
\usepackage[dvips]{epsfig}
\usepackage[usenames]{color}
\usepackage{latexsym}
\usepackage{epsfig}
\usepackage{amsmath}
\def\bslantfrac#1#2{{#1}\backslash\kern-0.1em{#2}}

\advance\topmargin by 2 cm

\begin{document}
  \newcommand{\Real}{\Re\text{e}}
  \newcommand{\Imag}{\Im\text{m}}
  \graphicspath{{FigureDraft/}}
  \title{Exclusive Neutral Pion Electroproduction in the Deeply Virtual Regime}
  \author{E.~Fuchey}
  \affiliation{Clermont Universit\'e, Universit\'e Blaise Pascal, CNRS/IN2P3, Laboratoire de Physique Corpusculaire, FR-63000 Clermont-Ferrand, France}
  \affiliation{Temple University, Philadelphia, Pennsylvania 19122, USA}
  \author{A.~Camsonne}
  \affiliation{Clermont Universit\'e, Universit\'e Blaise Pascal, CNRS/IN2P3, Laboratoire de Physique Corpusculaire, FR-63000 Clermont-Ferrand, France}
  \affiliation{Thomas Jefferson National Accelerator Facility, Newport News, Virginia 23606, USA}
  \author{C.~Mu\~noz~Camacho}
  \affiliation{CEA Saclay, IRFU/SPhN, FR-91191 Gif-sur-Yvette, France}
  \affiliation{Clermont Universit\'e, Universit\'e Blaise Pascal, CNRS/IN2P3, Laboratoire de Physique Corpusculaire, FR-63000 Clermont-Ferrand, France}
  \author{M.~Mazouz}
  \affiliation{LPSC, Universit\'e Joseph Fourier, CNRS/IN2P3, INPG, FR-38026 Grenoble, France}
  \affiliation{Facult\'e des sciences de Monastir, TN-5000 Tunisia}
  \author{G.~Gavalian}
  \affiliation{Old Dominion University, Norfolk, Virginia 23508, USA}
  \author{E.~Kuchina}
  \affiliation{Rutgers, The State University of New Jersey, Piscataway, New Jersey 08854, USA}
  \author{M.~Amarian}
  \affiliation{Old Dominion University, Norfolk, Virginia 23508, USA}
  \author{K.~A.~Aniol}
  \affiliation{California State University, Los Angeles, Los Angeles, California 90032, USA}
  \author{M.~Beaumel}
  \affiliation{CEA Saclay, IRFU/SPhN, FR-91191 Gif-sur-Yvette, France}
  \author{H.~Benaoum}
  \affiliation{Syracuse University, Syracuse, New York 13244, USA}
  \author{P.~Bertin}
  \affiliation{Clermont Universit\'e, Universit\'e Blaise Pascal, CNRS/IN2P3, Laboratoire de Physique Corpusculaire, FR-63000 Clermont-Ferrand, France}
  \affiliation{Thomas Jefferson National Accelerator Facility, Newport News, Virginia 23606, USA}
  \author{M.~Brossard}
  \affiliation{Clermont Universit\'e, Universit\'e Blaise Pascal, CNRS/IN2P3, Laboratoire de Physique Corpusculaire, FR-63000 Clermont-Ferrand, France}
  \author{M.~Canan}
  \affiliation{Old Dominion University, Norfolk, Virginia 23508, USA}
  \author{J.-P.~Chen}
  \affiliation{Thomas Jefferson National Accelerator Facility, Newport News, Virginia 23606, USA}
  \author{E.~Chudakov}
  \affiliation{Thomas Jefferson National Accelerator Facility, Newport News, Virginia 23606, USA}
  \author{B.~Craver}
  \affiliation{University of Virginia, Charlottesville, Virginia 22904, USA}
  \author{F.~Cusanno}
  \affiliation{INFN/Sezione Sanit\`{a}, IT-00161 Roma, Italy}
  \author{C.W.~de~Jager}
  \affiliation{Thomas Jefferson National Accelerator Facility, Newport News, Virginia 23606, USA}
  \author{A.~Deur}
  \affiliation{Thomas Jefferson National Accelerator Facility, Newport News, Virginia 23606, USA}
  \author{C.~Ferdi}
  \affiliation{Clermont Universit\'e, Universit\'e Blaise Pascal, CNRS/IN2P3, Laboratoire de Physique Corpusculaire, FR-63000 Clermont-Ferrand, France}
  \author{R.~Feuerbach}
  \affiliation{Thomas Jefferson National Accelerator Facility, Newport News, Virginia 23606, USA}
  \author{J.-M.~Fieschi}
  \affiliation{Clermont Universit\'e, Universit\'e Blaise Pascal, CNRS/IN2P3, Laboratoire de Physique Corpusculaire, FR-63000 Clermont-Ferrand, France}
  \author{S.~Frullani}
  \affiliation{INFN/Sezione Sanit\`{a}, IT-00161 Roma, Italy}
  \author{M.~Gar\c con}
  \affiliation{CEA Saclay, IRFU/SPhN, FR-91191 Gif-sur-Yvette, France}
  \author{F.~Garibaldi}
  \affiliation{INFN/Sezione Sanit\`{a}, IT-00161 Roma, Italy}
  \author{O.~Gayou}
  \affiliation{Massachusetts Institute of Technology, Cambridge, Massachusetts 02139, USA}
  \author{R.~Gilman}
  \affiliation{Rutgers, The State University of New Jersey, Piscataway, New Jersey 08854, USA}
  \author{J.~Gomez}
  \affiliation{Thomas Jefferson National Accelerator Facility, Newport News, Virginia 23606, USA}
  \author{P.~Gueye}
  \affiliation{Hampton University, Hampton, Virginia 23668, USA}
  \author{P.A.M.~Guichon}
  \affiliation{CEA Saclay, IRFU/SPhN, FR-91191 Gif-sur-Yvette, France}
  \author{B.~Guillon}
  \affiliation{LPSC, Universit\'e Joseph Fourier, CNRS/IN2P3, INPG, FR-38026 Grenoble, France}
  \author{O.~Hansen}
  \affiliation{Thomas Jefferson National Accelerator Facility, Newport News, Virginia 23606, USA}
  \author{D.~Hayes}
  \affiliation{Old Dominion University, Norfolk, Virginia 23508, USA}
  \author{D.W.~Higinbotham}
  \affiliation{Thomas Jefferson National Accelerator Facility, Newport News, Virginia 23606, USA}
  \author{T.~Holmstrom}
  \affiliation{College of William and Mary, Williamsburg, Virginia 23187, USA}
  \author{C.E.~Hyde}
  \affiliation{Old Dominion University, Norfolk, Virginia 23508, USA}
  \affiliation{Clermont Universit\'e, Universit\'e Blaise Pascal, CNRS/IN2P3, Laboratoire de Physique Corpusculaire, FR-63000 Clermont-Ferrand, France}
  \author{H.~Ibrahim}
  \affiliation{Old Dominion University, Norfolk, Virginia 23508, USA}
  \author{R.~Igarashi}
  \affiliation{University of Saskatchewan, Saskatchewan, SK, Canada, S7N 5C6}
  \author{F.~Itard}
  \affiliation{Clermont Universit\'e, Universit\'e Blaise Pascal, CNRS/IN2P3, Laboratoire de Physique Corpusculaire, FR-63000 Clermont-Ferrand, France}
  \author{X.~Jiang}
  \affiliation{Rutgers, The State University of New Jersey, Piscataway, New Jersey 08854, USA}
  \author{H.S.~Jo}
  \affiliation{Institut de Physique Nucl\'eaire CNRS-IN2P3, Orsay, France}
  \author{L.J.~Kaufman}
  \affiliation{University of Massachusetts Amherst, Amherst, Massachusetts 01003, USA}
  \author{A.~Kelleher}
  \affiliation{College of William and Mary, Williamsburg, Virginia 23187, USA}
  \author{A.~Kolarkar}
  \affiliation{University of Kentucky, Lexington, Kentucky 40506, USA}
  \author{G.~Kumbartzki}
  \affiliation{Rutgers, The State University of New Jersey, Piscataway, New Jersey 08854, USA}
  \author{G.~Laveissiere}
  \affiliation{Clermont Universit\'e, Universit\'e Blaise Pascal, CNRS/IN2P3, Laboratoire de Physique Corpusculaire, FR-63000 Clermont-Ferrand, France}
  \author{J.J.~LeRose}
  \affiliation{Thomas Jefferson National Accelerator Facility, Newport News, Virginia 23606, USA}
  \author{R.~Lindgren}
  \affiliation{University of Virginia, Charlottesville, Virginia 22904, USA}
  \author{N.~Liyanage}
  \affiliation{University of Virginia, Charlottesville, Virginia 22904, USA}
  \author{H.-J.~Lu}
  \affiliation{Department of Modern Physics, University of Science and Technology of China, Hefei 230026, China}
  \author{D.J.~Margaziotis}
  \affiliation{California State University, Los Angeles, Los Angeles, California 90032, USA}
  \author{Z.-E.~Meziani}
  \affiliation{Temple University, Philadelphia, Pennsylvania 19122, USA}
  \author{K.~McCormick}
  \affiliation{Rutgers, The State University of New Jersey, Piscataway, New Jersey 08854, USA}
  \author{R.~Michaels}
  \affiliation{Thomas Jefferson National Accelerator Facility, Newport News, Virginia 23606, USA}
  \author{B.~Michel}
  \affiliation{Clermont Universit\'e, Universit\'e Blaise Pascal, CNRS/IN2P3, Laboratoire de Physique Corpusculaire, FR-63000 Clermont-Ferrand, France}
  \author{B.~Moffit}
  \affiliation{College of William and Mary, Williamsburg, Virginia 23187, USA}
  \author{P.~Monaghan}
  \affiliation{Massachusetts Institute of Technology, Cambridge, Massachusetts 02139, USA}
  \author{S.~Nanda}
  \affiliation{Thomas Jefferson National Accelerator Facility, Newport News, Virginia 23606, USA}
  \author{V.~Nelyubin}
  \affiliation{University of Virginia, Charlottesville, Virginia 22904, USA}
  \author{M.~Potokar}
  \affiliation{Institut Jozef Stefan, University of Ljubljana, Ljubljana, Slovenia}
  \author{Y.~Qiang}
  \affiliation{Massachusetts Institute of Technology, Cambridge, Massachusetts 02139, USA}
  \author{R.D.~Ransome}
  \affiliation{Rutgers, The State University of New Jersey, Piscataway, New Jersey 08854, USA}
  \author{J.-S.~R\'eal}
  \affiliation{LPSC, Universit\'e Joseph Fourier, CNRS/IN2P3, INPG, FR-38026 Grenoble, France}
  \author{B.~Reitz}
  \affiliation{Thomas Jefferson National Accelerator Facility, Newport News, Virginia 23606, USA}
  \author{Y.~Roblin}
  \affiliation{Thomas Jefferson National Accelerator Facility, Newport News, Virginia 23606, USA}
  \author{J.~Roche}
  \affiliation{Thomas Jefferson National Accelerator Facility, Newport News, Virginia 23606, USA}
  \author{F.~Sabati\'e}
  \affiliation{CEA Saclay, IRFU/SPhN, FR-91191 Gif-sur-Yvette, France}
  \author{A.~Saha}
  \affiliation{Thomas Jefferson National Accelerator Facility, Newport News, Virginia 23606, USA}
  \author{S.~Sirca}
  \affiliation{Institut Jozef Stefan, University of Ljubljana, Ljubljana, Slovenia}
  \author{K.~Slifer}
  \affiliation{University of Virginia, Charlottesville, Virginia 22904, USA}
  \author{P.~Solvignon}
  \affiliation{Temple University, Philadelphia, Pennsylvania 19122, USA}
  \author{R.~Subedi}
  \affiliation{Kent State University, Kent, Ohio 44242, USA}
  \author{V.~Sulkosky}
  \affiliation{College of William and Mary, Williamsburg, Virginia 23187, USA}
  \author{P.E.~Ulmer}
  \affiliation{Old Dominion University, Norfolk, Virginia 23508, USA}
  \author{E.~Voutier}
  \affiliation{LPSC, Universit\'e Joseph Fourier, CNRS/IN2P3, INPG, FR-38026 Grenoble, France}
  \author{K.~Wang}
  \affiliation{University of Virginia, Charlottesville, Virginia 22904, USA}
  \author{L.B.~Weinstein}
  \affiliation{Old Dominion University, Norfolk, Virginia 23508, USA}
  \author{B.~Wojtsekhowski}
  \affiliation{Thomas Jefferson National Accelerator Facility, Newport News, Virginia 23606, USA}
  \author{X.~Zheng}
  \affiliation{Argonne National Laboratory, Argonne, Illinois 60439, USA}
  \author{L.~Zhu}
  \affiliation{University of Illinois, Urbana, Illinois 61801, USA}
  \collaboration{The Jefferson Lab Hall A Collaboration}
  
  
  
  \begin{abstract}
    We present measurements of the $ep\rightarrow ep\pi^0$ cross section extracted at two values of four-momentum transfer $Q^2=1.9 \; {\rm GeV}^2$ and $Q^2=2.3 \; {\rm GeV}^2$ at Jefferson Lab Hall A. 
    The kinematic range allows one to study the evolution of the extracted cross section as a function of $Q^2$ and $W$.
    Results 
    are confronted with Regge-inspired calculations and GPD predictions.
    An intepretation of our data within the framework of semi-inclusive deep inelastic scattering 
    is also discussed.
  \end{abstract}
  
  

  \maketitle
  
  \section{Introduction}
  \label{sec1}
  
\begin{figure}
    \centering
    \includegraphics[width=0.5\linewidth]{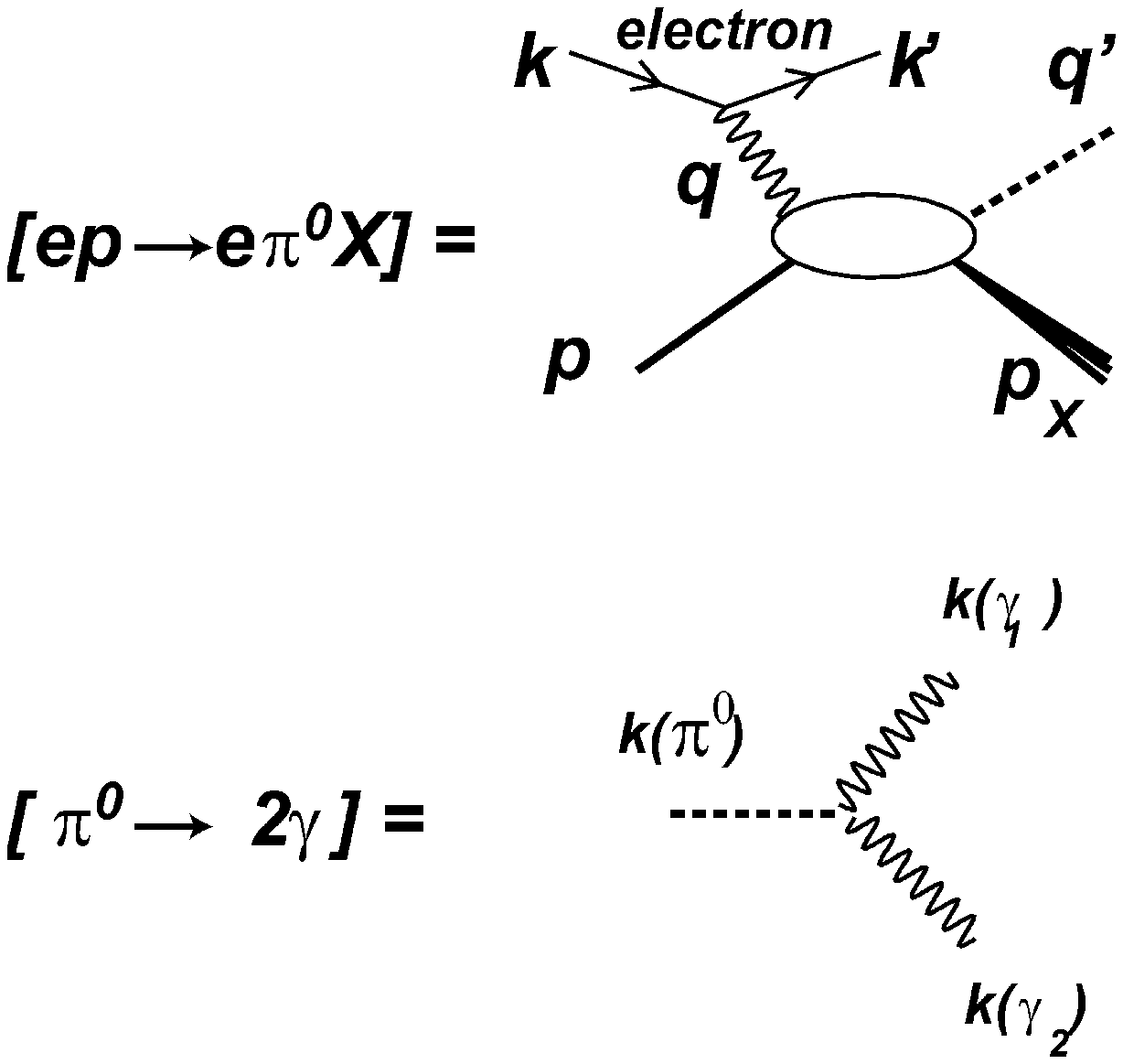}
    \caption{Diagram of the forward $\pi^0$ electroproduction reaction (top), and of the dominant $\pi^0$ decay mode (bottom). The kinematic invariants of this reaction are defined as: $Q^2 = -(k-k^{\prime})^2$, $x_{\rm Bj} = Q^2/(2 p  q)$, $t = (q-q^{\prime})^2$,   
    $W^2  = s=M_p + Q^2(1/x_{\rm Bj}-1)$, and 
    $t_{\rm min} = \frac{(Q^2-m_{\pi}^2)^2}{4s}-(|q^{\rm c.m.}|-|q^{\prime CM}|)^2$,
     with $|q^{\rm c.m.}|$ and $|q^{\prime CM}|$ the norms of $\vec{q}$, $\vec{q^{\prime}}$ in the 
     $p\pi^0$ final state center-of-mass frame.}
    \label{pi0Diag}
  \end{figure}

  The past decade has shown a strong evolution of the study of hadron structure through exclusive processes, allowing access to the three-dimensional structure of  hadrons.
  Exclusive processes include deeply virtual Compton scattering (DVCS) and deeply virtual meson production (DVMP).
  This document focuses on the latter, and more precisely on neutral pion production.
  \\
  We present measurements of the differential cross section for the forward exclusive electroproduction reaction $ep\,\rightarrow\,ep\pi^0$, through virtual photoabsorption. 
  A diagram of this process, including definitions of the kinematic variables, is presented in Figure \ref{pi0Diag}.
  Results will be presented for four kinematics. 
  Two of them are defined by the same value of $x_{\rm Bj} = 0.36$ and are called 
  Kin2 (at $Q^2 = 1.9 \; {\rm GeV}^2$) and Kin3 (at $Q^2 = 2.3 \; {\rm GeV}^2$)
  The two remaining ones are defined by the same value of $Q^2 = 2.1 \; {\rm GeV}^2$ and are called 
  Kin$X$2 (at $x_{\rm Bj} = 0.40$) and Kin$X$3 (at $x_{\rm Bj} = 0.33$).
  
  The behavior of the cross section will be compared to different models that are available to describe $\pi^0$ electroproduction, including the Regge model and the generalized parton distribution (GPD) framework. 
  
  Forward photo-production at asymptotically high energies can be described by the Regge theory, which exploits the analytic properties of the scattering amplitude in the limit $t/s \rightarrow 0$ \cite{Goldstein:1973xn}. 
  Previous analyses have applied Regge phenomenology to exclusive photo- and electro-production in the kinematic range presented here \cite{VGL_NPA, VGL_PLB}.
  Recent computations with Regge-inspired models exist for our kinematics.
  These models include $\rho$, $\omega$, and $b$ meson exchange as well as $\pi^{\pm}$ rescattering. 
  Among these, there is the $t$-channel meson-exchange (TME) model by Laget {\it et al}. 
  A brief description of this model has been given in \cite{pi0BSA}, and it is described extensively in \cite{VGL_PRC, LagetNew}.
  Recent JLab Hall C experiments studying the $Q^2$ dependence of charged-pion electroproduction with a longitudinal-transverse separation were analyzed using the TME formalism \cite{HallCpiplus_Long}.
  Another Regge-inspired computation by Ahmad, Goldstein and Liuti \cite{Simoneta} is available for our kinematics.
  
  In the Bjorken limit $Q^2 \rightarrow \infty$, and $t/Q^2 \ll 1$ at fixed $x_{\rm Bj}$, 
  the scattering amplitude is dominated by the leading order (or leading twist) amplitude of GPDs and the pion distribution amplitude (DA) \cite{Fact_HXE, VGP, VGG}. 
  The GPDs are light-cone matrix elements of non-local bilinear quark and gluon operators \cite{FP42, JI_PRL78, NonForward}, unifying the elastic electroweak form factors with the forward parton distributions of deep-inelastic lepton scattering. 
  Cross section predictions within the GPD framework exist for the longitudinal cross section $\sigma_L$ \cite{VGP, VGG}.
  With the definitions of \cite{Fact_HXE, VGP, VGG}, the cross sections are predicted to scale as $\sigma_L\sim Q^{-6}$ and $\sigma_T\sim Q^{-8}$.  
  Thus at sufficiently high $Q^2$, $\sigma_L$ will dominate over $\sigma_T$.
  Beam spin asymmetries for forward exclusive $\pi^0$ electroproduction have been measured for $Q^2>1 \; \rm{GeV}^2$ \cite{pi0BSA}.
  We performed measurements at two $Q^2$ values at fixed $x_B$ in order to test these predictions of $Q^2$ dependence.
  An interpretation of exclusive data with semi-inclusive mechanisms also exists to explain transverse cross sections of hard exclusive charged-pion electroproduction \cite{Lund}.

  In the second section details of the experiment are presented, while the third section is devoted to the calibration of the calorimeter.
  The formalism of $\pi^0$ electroproduction by Drechsel and Tiator \cite{DT} is presented in the fourth section with a special emphasis on the expressions for the hadronic tensors.
  The fifth section is devoted to the extraction of the cross sections and the sixth and seventh sections to the radiative corrections and the evaluation of the systematic errors.
  Finally, our results are presented in Sec. \ref{sec8}, with a discussion and conclusions in Secs. \ref{sec9} and \ref{sec10}, respectively.
  
  \section{Experiment}
  \label{sec2}
  
\begin{figure}
    \centering
    \includegraphics[width=0.9\linewidth]{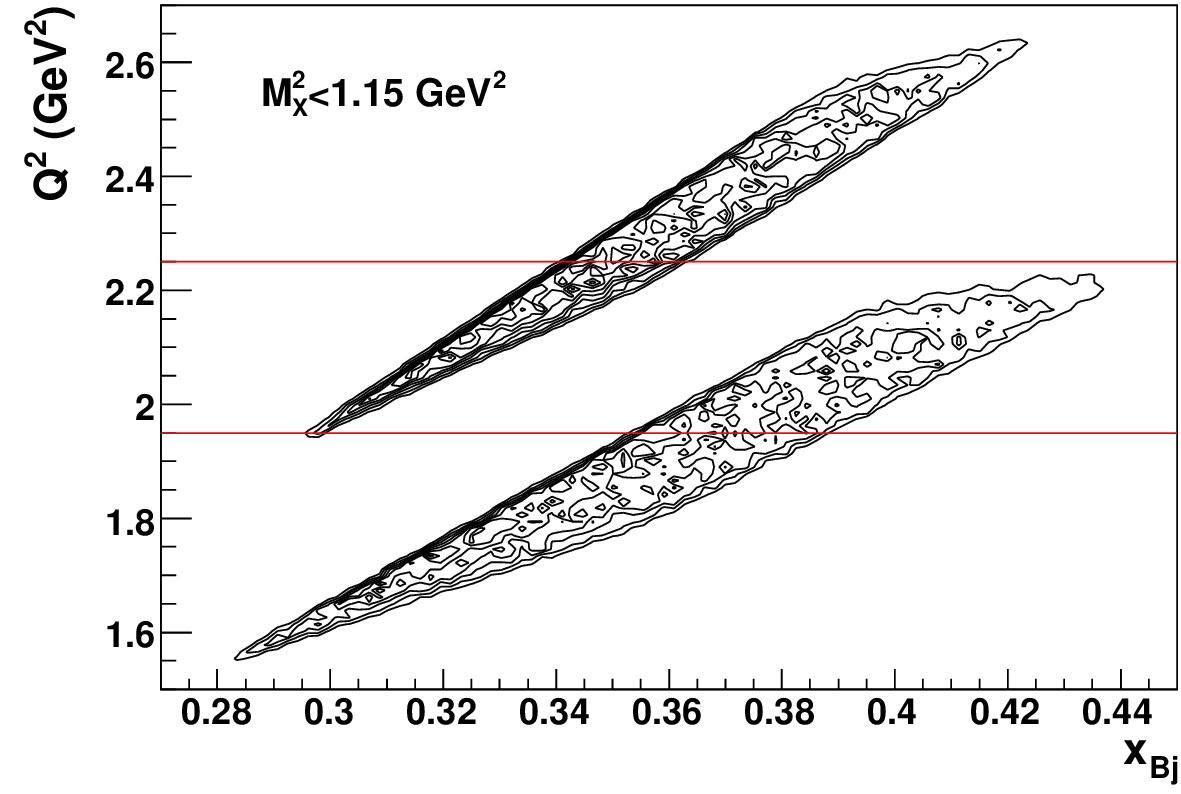}
    \caption{(Color online) Distribution of $H(e,e^{\prime}\pi^0)X$ events in the [$x_{\rm Bj}$, $Q^2$] plane, for Kin2 ($x_{\rm Bj}=0.36$, $Q^2=1.9\; {\rm GeV}^2$) and Kin3 ($x_{\rm Bj}=0.36$, $Q^2=2.3\; {\rm GeV}^2$).
    Events for Kin$X$2 ($x_{\rm Bj}=0.40$, $Q^2=2.1\; {\rm GeV}^2$) and Kin$X$3 ($x_{\rm Bj}=0.33$, $Q^2=2.1\; {\rm GeV}^2$)  are bounded by the two horizontal lines.}
    \label{Q2vsxB}
  \end{figure}

  The present data were acquired as part of Jefferson Lab Hall A experiment E00-110 \cite{DVCSProposal}.
  Additional details about the experimental configuration, calibrations, and analysis can be found in \cite{CarlosThese, CarlosPRL}.
  This paper reports on the analysis of the triple coincidence $H(e,e^{\prime}\gamma\gamma)X$ events. 
  A 5.75 GeV electron beam was incident on a 15 cm liquid hydrogen target, for a typical luminosity of $10^{37} \; {\rm cm}^{-2} \, {\rm s}^{-1}$.
  Electrons were detected in a high resolution spectrometer (HRS).
  Photons were detected in a 132 element ${\rm PbF}_2$ calorimeter, each of the elements measuring $3 \times 3 \,{\rm cm}^2 \times 20 X_0$. 
  The high resolution allows one to accurately define (1) the virtual photon, having the kinematics centered at a fixed $x_{\rm Bj} = 0.36$ and two values of $Q^2 = 1.9$ and $2.3 \; {\rm GeV}^2$, as shown in Figure \ref{Q2vsxB} and (2) the real photon momentum unit vector, thanks to the vertex resolution of the HRS, and the position resolution of the electromagnetic calorimeter.
  The validation threshold for the data acquisition trigger was set to about 1 GeV for each photon cluster. 
  For the exclusive $\pi^0 \rightarrow \gamma\gamma$ events, the minimum distance between the centroids of the two clusters that guarantees separation is about 10 cm. 
  This is achieved by the minimal opening angle $\approx 2 m_{\pi}/E_{\pi}$ and the distance from the center of the target to the calorimeter front face $L=110$\,cm.
  The achieved coincidence resolving time between the scattered electron and either photon cluster is 0.6 ns, rms.
  
  Figure \ref{Q2vsxB} shows the distribution of $H(e,e^{\prime}\pi^0)X$ events in the [$x_{\rm Bj}$, $Q^2$] plane, for missing mass squared $M_X^2 = (q+p-q^{\prime})^2 \leq 1.15 \; {\rm GeV}^2$.
  The analysis relies only on two specific qualities of the experiment:
  \begin{enumerate}
  \item{Thanks to the resolution of the spectrometer and the calorimeter, one can use the missing-mass squared to ensure exclusivity. The exclusive sample is selected by putting a cut on the missing-mass squared at the proton plus the pion mass squared.}
  \item{For exclusive events, the reconstruction of the invariant momentum transfer $t$ and $t_{\rm min}$ relies on the positions of the reconstructed photons, of which the resolution is better than that of the energy. From this, a resolution in $t$ better than that in the energy is obtained. All data are presented as a function of $t_{\rm min}-t$, which is directly linked to the angle of the pion production relative to the virtual photon direction in the center of mass $\theta_{\pi}^{\rm c.m.}$: $t_{\rm min}-t = 2q^{\rm c.m.}q^{\prime CM}(1-\cos{\theta_{\pi}^{\rm c.m.}})$.}
  \end{enumerate}
  
\begin{figure}
    \centering
    \includegraphics[width=0.9\linewidth]{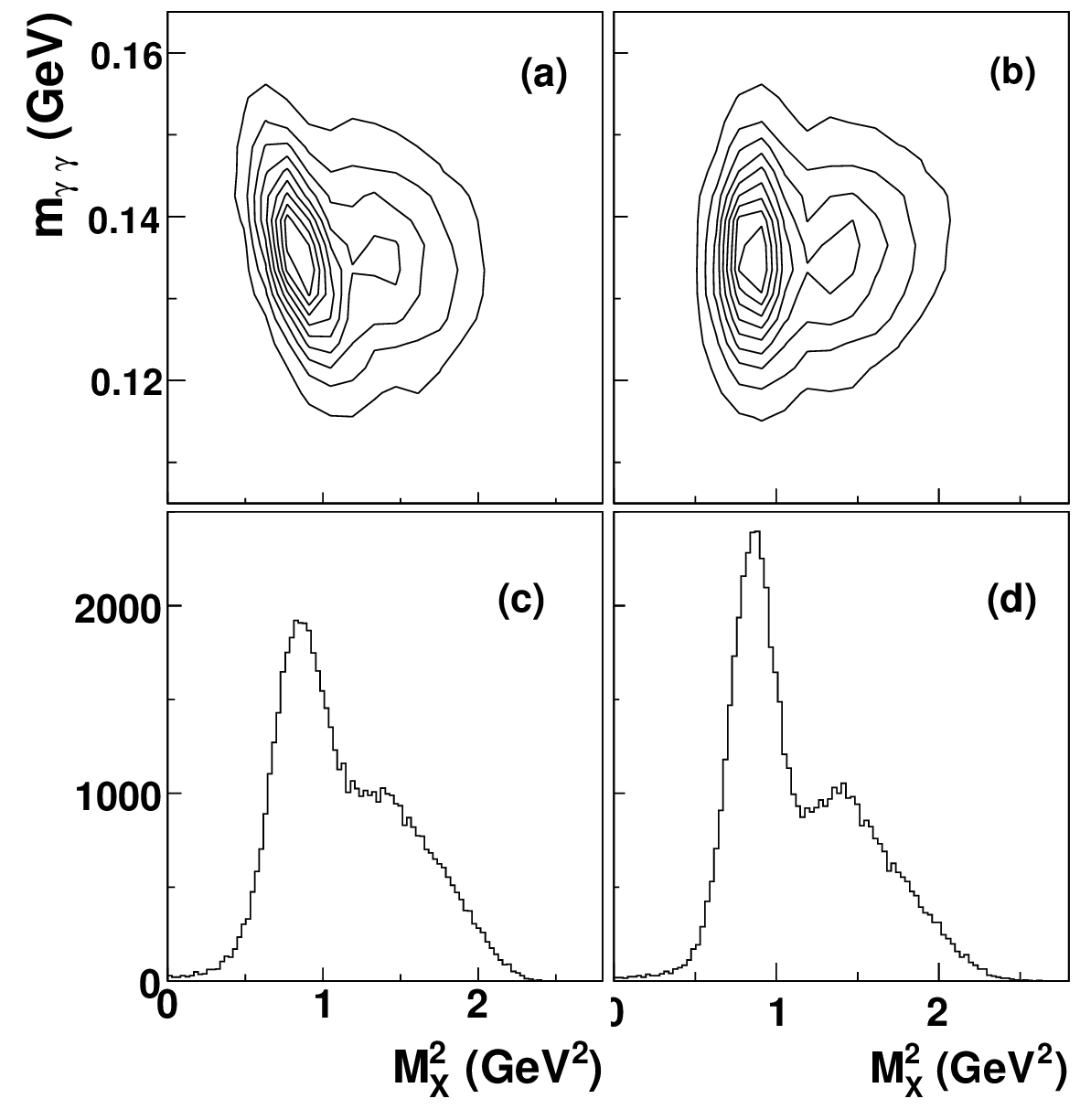}
    \caption{[(a),(b)] Distributions of $H(e,e^{\prime}\gamma\gamma)X$ events within cuts in the [$M_X^2$, $m_{\gamma\gamma}$] plane for Kin3. (a) Raw distribution showing a clear correlation between these two variables. (b) The same distribution after a rotation around ($M_p^2$, $m_{\pi^0}$) to improve the $M_X^2$ resolution. [(c),(d)] Projections on the $M_X^2$ axis of the [$M_X^2$, $m_{\gamma\gamma}$] distributions shown, respectively, in (a) and (b). The lower right panel shows that the resolution is indeed improved by the rotation.}
    \label{MX2rotation}
  \end{figure}

  In the $ep \rightarrow e^{\prime}\gamma_1 \gamma_2 X$ reaction, there are six four-vectors, equivalent to 24 independent kinematic variables.
  The measured four-vectors $k$, $p$, and $k^{\prime}$, and four-momentum conservation, reduce the number of independent variables to eight. 
  The measurement of the two directional vectors $\hat{k}(\gamma_1) = \vec{q_1}/q_1$ and $\hat{k}(\gamma_2) = \vec{q_2}/q_2$ from the target vertex (reconstructed by the HRS) to the two cluster positions in the calorimeter provides four more kinematic constraints. 
  Finally, the hypothesis that the observed calorimeter showers are due to photons  ($m_{q_1} = m_{q_2} = 0$) provides two more kinematic constraints.
  The remaining two unknowns, which we express as $m_{\gamma\gamma}^2 = (q_1+q_2)^2$ and $M_X^2$, are determined by the previous constraints plus the energy of the two photons. 
  Figure \ref{MX2rotation} displays the distribution of the $H(e,e^{\prime}\gamma\gamma)X$ events in the [$M_X^2$, $m_{\gamma\gamma}$] plane, for Kin3.
  The upper left panel of this figure shows a clear correlation between the two variables in the exclusive region ($M_X^2 \simeq M_p^2$). 
  This is a consequence of resolution fluctuations in the energies $E_1$ and $E_2$ of the two photons issued from a $\pi^0$, which correlate fluctuations in $M_X^2$ and $m_{\gamma\gamma}$.
  The missing mass in the right-hand panels is obtained by an empirical adjustment:
  \begin{equation}
    \left. M_X^2 \right|_{\rm corr} = \left. M_X^2 \right|_{\rm raw}+ C (m_{\gamma \gamma} - m_{\pi}),
  \end{equation}
  with $C = 13 \; {\rm GeV}$.
  
  This transformation produces a noticeable improvement in the $M_X^2$ distribution (lower right panel of Figure \ref{MX2rotation}).
  
  \section{Calibration}
  \label{sec3}
  
  We performed elastic $H(e,e_{\rm calo}^{\prime} \; p_{\rm HRS})$ calibrations at the beginning, middle, and end of the experiment \cite{Calib}. 
 The calorimeter was retracted to a position at 5.5 m from the target, in order to optimize the electron coverage in the calorimeter with the proton acceptance of the HRS. 
  These data were used for the block calibration.
  After calibration the calorimeter energy resolution was observed to be 2.4\% at 4.2 GeV with a position resolution of 2 mm at 110 cm from the target. 
  The elastic data also provided a consistency check on the efficiency of the detectors and all associated electronics from the observation that the elastic cross section agreed with the Kelly form-factor  parametrization \cite{Kelly} at the 1.1\% level.
  During the experiment, the light output from the ${\rm PbF}_2$ blocks decreased by up to 20\%, strongly correlated with the distance of the blocks from the beam line. 
  We attribute this to radiation damage of the blocks. In addition, seven blocks, at random positions, showed much higher radiation damage. 
  One explanation could be a poorer crystal quality of those crystals. 
  We adjusted the calibration of each block, assuming an independent linear dose {\it versus} attenuation curve. 
  In addition to radiation damage, each crystal received a pileup of low-energy photons in random coincidence,
 resulting in a degradation of the energy resolution, and in a shift in the calibration as a function of its distance to the beam line.
  This effect was taken into account through successive steps:
  \begin{enumerate}
  \item{For each block the position of the reconstructed missing-mass squared peak was centered at $M_p^2$ through an energy calibration of the experimental data.}
  \item{A {\sc geant} simulation generated a sharper resolution in missing mass than the experimental data for each calorimeter block. For each block the energy of the simulation was calibrated together with a simultaneous energy smearing, in order to center the reconstructed missing-mass peak position at $M_p^2$, and to equate the resolution of the simulation to that of the experimental data.}
  \end{enumerate}
  These calibrations are explained in the following paragraphs.
  
\begin{figure}
    \centering
    \includegraphics[width=0.5\linewidth]{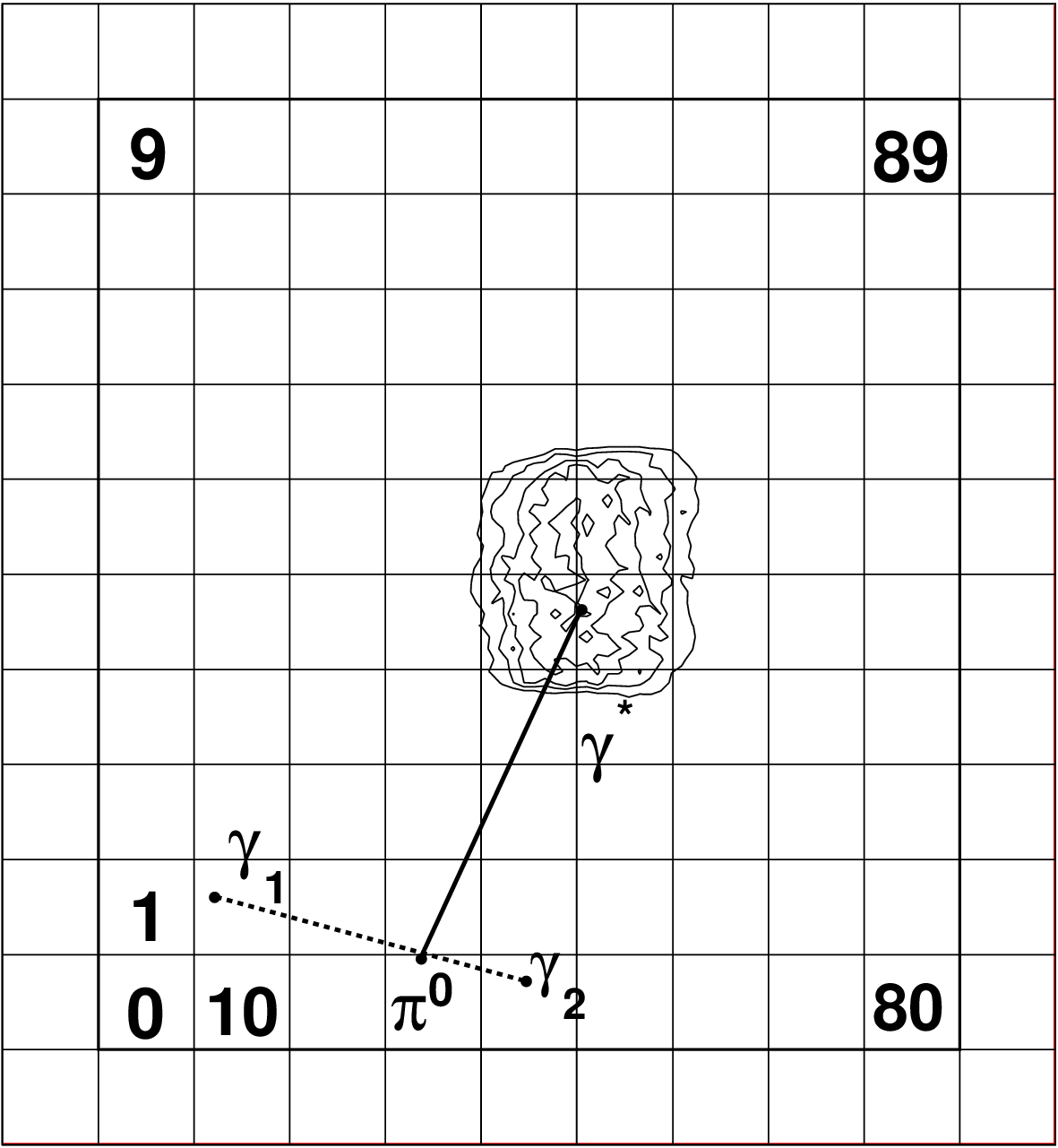}
    \caption{Projection on the calorimeter of the virtual photons $\gamma^*$ within cuts for Kin3. Also shown is the block relabeling used for the calorimeter calibration described in the text. The calorimeter is viewed from the rear, with the downstream beam passing to the right.}
    \label{CaloProj}
  \end{figure}

  We consider only the 90 blocks of the inner calorimeter (see Figure \ref{CaloProj} for the labeling), indexed by $\mu$.
  We will assume that the energy of the photon is driven by the block where the shower makes the largest energy deposit.
  The 90 distributions of missing-mass squared $(M_X^2)_{\mu}^i = (k+P-k^{\prime}-q_{\mu}-q_{\nu})_i^2 = (E_X^2)^i - (\vec{P_X}^2)^i$ are built with all events $i$ involving block $\mu$.
  Note that for each event $i$, the reconstructed missing-mass squared appears in two distributions. 
  To compare these distributions, two estimators are constructed: the mean $\langle M_X^2 \rangle_{\mu}$ and the sigma $\sigma_{\mu}$ of a Gaussian fitted to these distributions, over a limited range ($0.62 \; {\rm GeV}^2 < (M_X^2)_{\mu} < 1.09 \; {\rm GeV}^2$). 
  The calorimeter is calibrated using
  \begin{equation}
    \Delta M_X^2 = -2 \Delta q_{\mu}\left(E_X - \frac{\vec{P_X}.\vec{q_{\mu}}}{|q_{\mu}|}\right),
  \end{equation}
  with $\Delta M_X^2 = \langle (M_X^2)_{\mu} \rangle - M_p^2$.
  Neglecting the $P_X$ term compared to $E_X$ between the parentheses, we obtain an energy correction:
  \begin{equation}
    q_{\mu}^i \rightarrow q_{\mu}^i + \Delta q_{\mu}^i = q_{\mu}^i + \frac{\Delta M_X^2}{2 (E_X)^i}.
    \label{datacalib}
  \end{equation}
  We recall here that each event involves two blocks. 
  The reconstructed missing mass of one block is then influenced by contributions from all other blocks. 
  Because of this, several iterations are necessary.
  Then, the missing-mass distribution of each block for simulated events is adjusted to get the same missing-mass position and resolution as the experimental missing-mass distribution.
  The missing-mass cut applied to ensure exclusivity is the same for simulation and data, and if the resolution is better for simulation, applying such a cut will remove more experimental events than simulation particularly near the beam where the noise degrades the experimental resolution.
  This gives a spurious contribution to the $\cos{\,\phi_{\pi}}$ term which has to be removed by smearing the simulation resolution.
  To this purpose, the momentum of each event $i$ at the $n{\rm th}$ iteration contributing to the $M_X^2$ distribution of the block $\mu$  is changed from $(\vec{q_{\mu}})_{n-1}^i$ to $(\vec{q_{\mu}})_n^i$ with a sampling from a Gaussian distribution:
  \begin{equation}
    (\vec{q_{\mu}})_n^i = \frac{(\vec{q_{\mu}})_{n-1}^i}{|q_{\mu}|_{n-1}^i} {\rm Gauss} \left( (q_{\mu})_{n}^{i}, \frac{\Delta \sigma_{\mu}}{\sqrt{2}} \right),
  \end{equation}
  where
  \begin{eqnarray}
    \Delta \sigma_{\mu} = \sqrt{(\sigma_{\mu})_{\rm data}^2-(\sigma_{\mu})_{\rm simu}^2},~~~~(\sigma_{\mu})_{\rm data} > (\sigma_{\mu})_{\rm simu}\\
    \Delta \sigma_{\mu} = 0,~~~~(\sigma_{\mu})_{\rm data} < (\sigma_{\mu})_{\rm simu}
  \end{eqnarray}
  and $(q_{\mu})_{n}^{i}$ is given by equation \eqref{datacalib}, except we put $\Delta M_X^2 = (\langle (M_X^2)_{\mu} \rangle_{simu}-\langle (M_X^2)_{\mu} \rangle_{data})/2$ in this case.
  The factor 2 in the denominator of $\Delta M_X^2$ is used to ensure a smooth convergence.

  \begin{table}[htbp]
    \begin{center}
      \caption[]{Mean deviation and resolution width of the $\pi^0\,\rightarrow\,\gamma \gamma$ reconstruction of the data and simulation. Events are selected by $M_{X}^{2}<1.15 \; {\rm GeV}^{2}$ and calorimeter threshold $E_{\rm thr} = 1.0$ GeV.}
      \label{mgg}
      \begin{tabular}{|l|c|c|}
	\hline
	\hline
	& $\langle m - m_{\pi^0} \rangle$ (GeV) & $\sqrt{\langle (m - m_{\pi^0})^2 \rangle}$ (GeV) \\
	\hline
	\multicolumn{3}{|c|}{Kin3}\\
	\hline
	Data & $-$0.00081 & 0.0088 \\
	\hline
	Simulation & +0.00072 & 0.0089 \\
	\hline
	\multicolumn{3}{|c|}{Kin2}\\
	\hline
	Data & $-$0.00017 & 0.0079 \\
	\hline
	Simulation & +0.00191 & 0.0085 \\
	\hline
	\hline
      \end{tabular}
    \end{center}
  \end{table}
  \begin{figure}
    \centering
    \includegraphics[width = 0.9\linewidth]{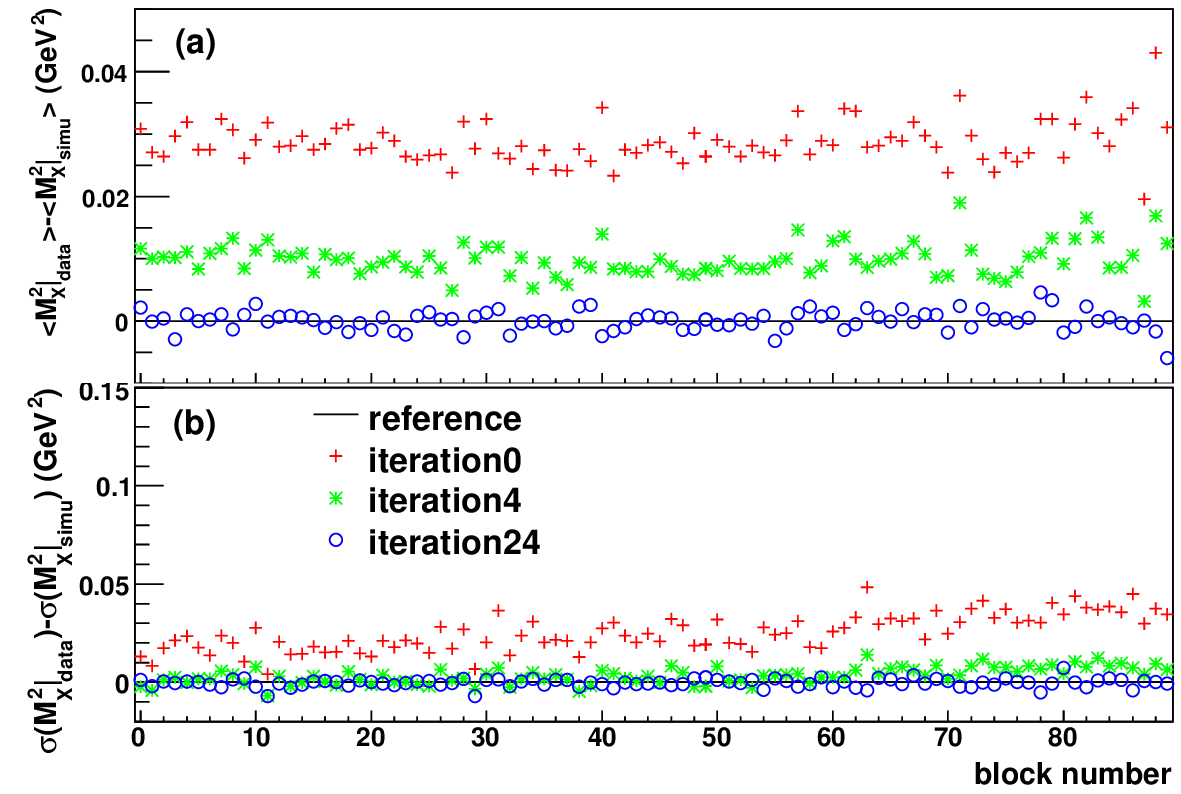}
    \caption{(Color online) Different iterations of the calibration for Kin2. The differences between simulation and data of the missing-mass peak position (a) and resolution (b) are shown before calibration (crosses), after calibration (open circles), and at a random iteration during calibration (asterisks).}
    \label{Calibration}
  \end{figure}
  \begin{figure}
    \centering
    \includegraphics[width = 0.9\linewidth]{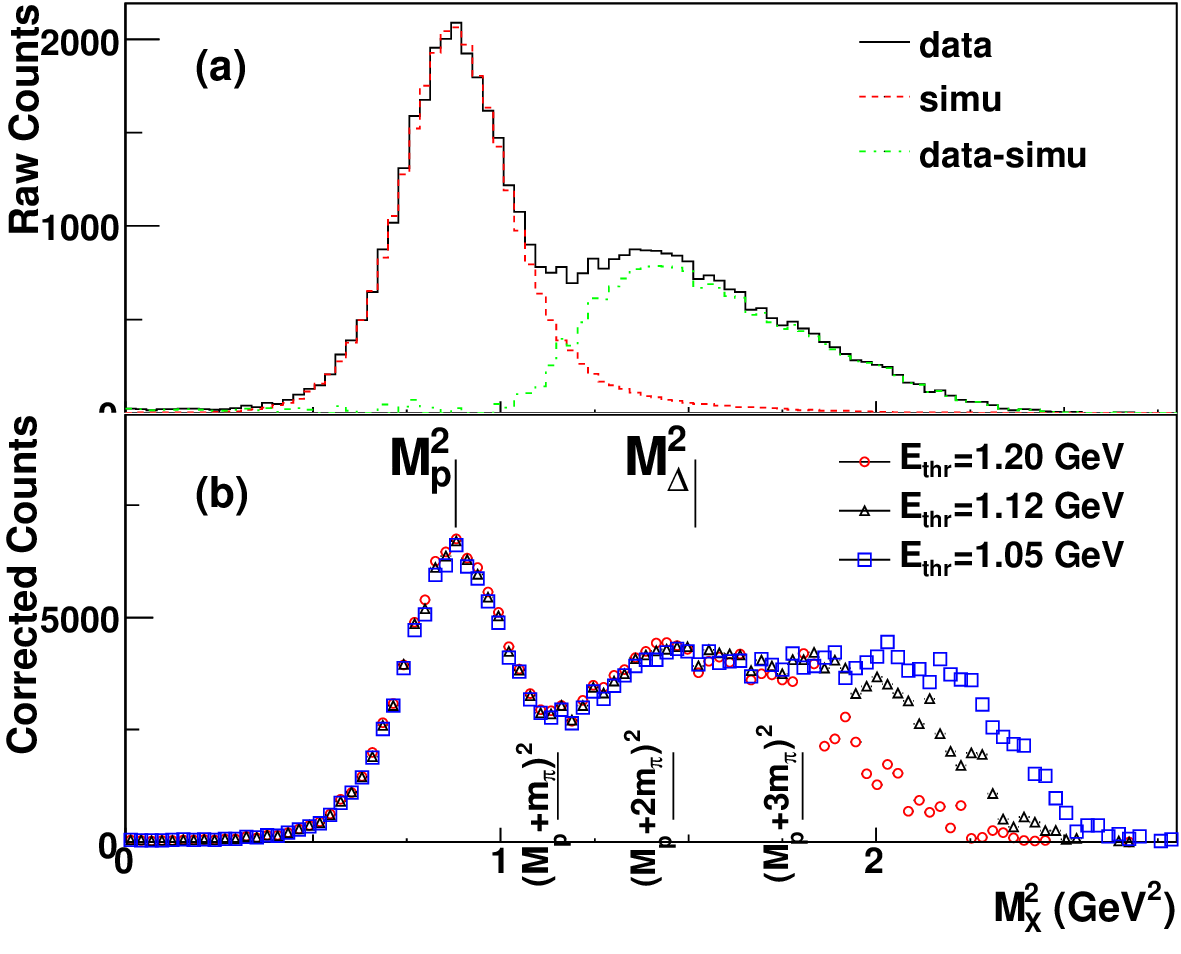}
    \caption{(Color online) (a) Raw $H(e,e^{\prime}\pi^0)X$ missing-mass distribution for Kin3 (solid histogram) compared to the simulation (dashed histogram), and the difference between the two (dotted histogram). (b) $H(e,e^{\prime}\pi^0)X$ missing-mass distribution at different values for the calorimeter threshold, corrected with a factor $1/(1-2(E_{\rm thr}/|\vec{p_{\pi}}|))$. This correction adds to the distribution all $\pi^0$ events missed because of the threshold value.}
    \label{MX2distrib}
  \end{figure}

  The results of these iterations are shown in Figure \ref{Calibration}.
  Figure \ref{MX2distrib} and Table \ref{mgg} illustrate the quality of the final calibration adjustments.
  The calibration of the missing-mass squared was cross-checked by comparing the invariant-mass distribution of both photons in each event. 
  Table \ref{mgg} lists the mean values of these distributions with respect to the pion mass, and their resolution.
  The agreement of the calibration with the data is at the 1.9 MeV level, while the widths of these distributions agree to better than 1 MeV.

  \section{Cross-section analysis}
  \label{sec4}
  
  In order to extract the differential cross section, it is advantageous to incorporate all model-independent kinematic dependences of the differential cross section into the experimental simulation.
  To this end, we express the differential cross section in terms of structure functions as described in the paper of Drechsel and Tiator \cite{DT} directly related to bilinear combinations of the Chew-Goldberger-Low-Nambu (CGLN) helicity amplitudes \cite{CGLN}.
  We define the differential phase-space elements $d^3 \Phi_e =  dQ^2 dx_{\rm Bj} d\phi_e$ and $d^5\Phi = d^3 \Phi_e d[t_{\rm min}-t] d\phi_{\pi}$ and the equivalent real photon energy in the c.m. frame $k_{\gamma}^{\rm c.m.} = (W^2-M_p^2)/2W$. Here $t_{\rm min} = \frac{(Q^2-m_{\pi}^2)^2}{4s}-(|q^{\rm c.m.}|-|q^{\prime {\rm c.m.}}|)^2$
with $|q^{\rm c.m.}|$ and $|q^{\prime {\rm c.m.}}|$ the norms of $\vec{q}$,
$\vec{q^{\prime}}$ in the center-of-mass frame.
  All these quantities are defined using the convention of Drechsel and Tiator \cite{DT}: $\hat{z}$ axis along the virtual photon, $\hat{y} = (\hat{k}_i \wedge \hat{k}_f)/ \sin{\theta_e}$ orthogonal to the leptonic plane, and $\hat{x} = \hat{y} \wedge \hat{z}$.

  To lowest order in the fine-structure constant $\alpha$, the differential cross section for an electron of helicity $h$ is
  \begin{equation}
    \frac{d^5 \sigma(h)}{d^5 \Phi} = \Gamma \frac{d^2 \sigma_v(h)}{dt d\phi_{\pi}},
  \end{equation}
  \begin{equation}
    \Gamma = \frac{\alpha}{2\pi^2}\frac{k^{\prime}}{k}\frac{k_{\gamma}}{Q^2}\frac{1}{1-\epsilon},
  \end{equation}
  with $k_{\gamma} = (W^2-M_p^2)/2M_p$ and $k$ and $k^{\prime}$ the energies of the incident and scattered electron, respectively.
  The virtual photo-absorption cross section is expanded as
  %
  \begin{align}
    \frac{d^2 \sigma_v(h)}{dt d\phi_{\pi}}&=\frac{1}{2q^{\rm c.m.}k_{\gamma}^{\rm c.m.}}\{ R_T + \epsilon_L R_L+\epsilon R_{TT}\cos{2\phi_{\pi}}
    \nonumber \\
    &+ \sqrt{2 \epsilon_L (1+\epsilon)} R_{TL} \cos{\phi_{\pi}}
    \nonumber \\
    &+ h\sqrt{2 \epsilon_L (1-\epsilon)} R_{TL^{\prime}} \sin{\phi_{\pi}} \}
  \label{dsigmavdt}
  \end{align}
  %
  where $q^{\rm c.m.} = |\vec{q}| \times M_p/W$ is the c.m. virtual photon three-momentum, $\epsilon = 1/[1+2(q^2/Q^2)\tan^2{\,\theta_e/2}]$ is the degree of linear polarization of the virtual photons, and $\epsilon_L/\epsilon = 4M_p^2x_{\rm Bj}^2/Q^2$.
  The response functions are defined as functions of the usual hadronic tensor $W^{\mu \nu}$:
  \begin{eqnarray}
    R_T = \frac{W_{xx}+W_{yy}}{2},\\
    R_L = W_{zz},\\
    \cos{\phi_{\pi}}R_{TL} = -{\rm Re} W_{xz},\\
    \sin{\phi_{\pi}}R_{TL^{\prime}} = -{\rm Im} W_{yz},\\
    \cos{2 \phi_{\pi}}R_{TT} = \frac{W_{xx}-W_{yy}}{2}.
  \end{eqnarray}
  
  The interference terms $R_{TL}$ and $R_{TL^{\prime}}$ have a leading $\sin{\,\theta_{\pi}^{\rm c.m.}}$ dependence, and the linear polarization interference term $R_{TT}$ has a leading $\sin^2{\,\theta_{\pi}^{\rm c.m.}}$ dependence.
  For this reason, we define reduced structure functions $r_{\Lambda}$, which remove this phase-space dependence, which are directly related to bilinear combinations of the CGLN helicity amplitudes $F_i$ \cite{CGLN}:
  \begin{equation}
    \begin{aligned}
      \left(\begin{array}{c}r_{TL}\\
	r_{TL^{\prime}}\end{array}\right) &= \frac{1}{\sin{\theta_{\pi}^{\rm c.m.}}} \left(\begin{array}{c}R_{TL}\\
	R_{TL^{\prime}}\end{array}\right),
    \end{aligned}
    \label{rTL_TLprim}
  \end{equation}
  \begin{equation}
    \begin{aligned}
	r_{TT} &= \frac{R_{TT}}{\sin^2{\theta_{\pi}^{\rm c.m.}}},
    \end{aligned}
    \label{rTT}
  \end{equation}
  \begin{equation}
    r_L = R_L,
    \label{rT}
  \end{equation}
  \begin{equation}
    r_T = R_T.
    \label{rL}
  \end{equation}

  Since our kinematics cover a wide range in $x_{\rm Bj}$ as well as in $Q^2$, we also have to include the $Q^2$ and the $W$ dependence of the  hadronic tensor $(W_{xx}+W_{yy})/2+\epsilon_L W_{zz} = r_T+\epsilon_L r_L$.
 We
 perform a preliminary extraction of the cross section on the kinematic points Kin2 and Kin3 (respectively Kin$X$2 and Kin$X$3) to get an estimate of the $Q^2$ (respectively, $W$) dependence of the hadronic tensor.
  The extracted $Q^2$ and $W$ dependences 
  are then 
  introduced explicitly in the formalism to perform a second ``definitive'' extraction.  The dependence is modeled in the
  form $(Q^2)^n$ and $W^\delta$.
  With the first iteration, the cross sections changed by $3\%$, but with a second iteration the cross sections changed by only $0.3\%$.
  %
  %
  
  The results will be presented as four separated cross sections following the usual decomposition found in the literature: 
  %
  \begin{align}
    \frac{d^2\sigma_v}{dt d\phi_{\pi}} &= \frac{1}{2 \pi} \left\{ \frac{d\sigma_T}{dt} + \epsilon_L \frac{d\sigma_L}{dt} \right.
    \nonumber \\
    &+ \sqrt{2 \epsilon_L (1+\epsilon)}\frac{d\sigma_{TL}}{dt} \cos{\phi_{\pi}} 
    \nonumber \\
    &+ \epsilon \frac{d\sigma_{TT}}{dt} \cos{2 \phi_{\pi}} 
    \nonumber \\
    &+ \left. h\sqrt{2 \epsilon_L (1-\epsilon)}\frac{d\sigma_{TL^{\prime}}}{dt} \sin{\phi_{\pi}} \right\}.    
    \label{dsigmav_fdsigmadt}
  \end{align}
  %
  
  \section{Extraction}
  \label{sec5}
  
  We define a compact notation that summarizes Eq. \eqref{dsigmavdt} in the form
  \begin{equation}
    \frac{d^5\sigma}{d^5\Phi} = \sum_{\Lambda} \frac{d^3 \Gamma_{\Lambda}}{d^3\Phi_e}r_{\Lambda} = \sum_{\Lambda} {\cal F}_{\Lambda}(x_v) r_{\Lambda}
    \label{dsigmavdt_compact}
  \end{equation}
  with ${\cal F}_{\Lambda}(x_v)$ containing all the kinematic dependence, $\Lambda \in \{ T+\epsilon_L L, TL, TT, TL^{\prime} \}$ and $x_v$ summarizing all variables $k, Q^2, x_{\rm Bj}, W, t$, considered at the vertex. 
  $T+\epsilon_L L$ reflects the fact that we used only one incident energy and consequently, we were not able to disentangle $d \sigma_T$ and $d \sigma_L$.
  This notation will be convenient to use for the presentation of the extraction process.

  \begin{figure}
    \centering
    \includegraphics[width = 0.9\linewidth]{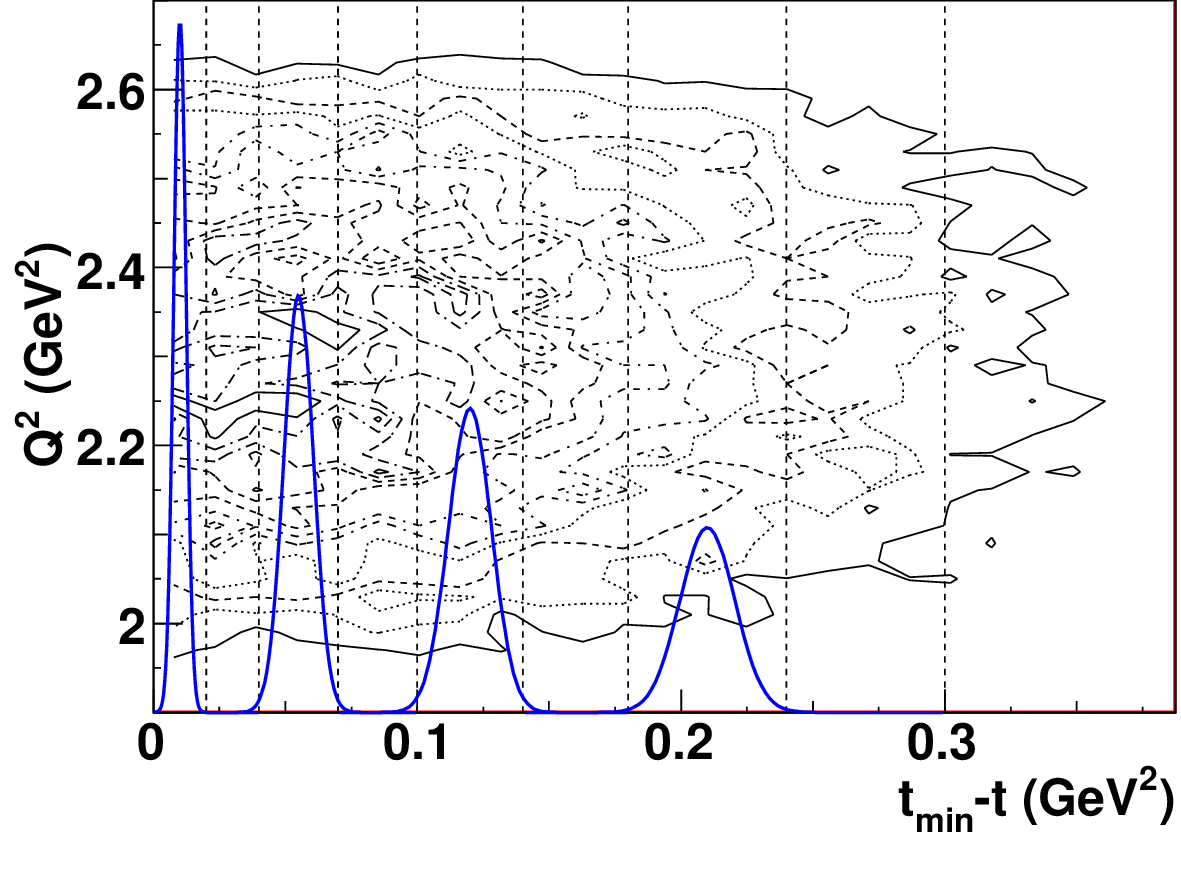}
    \caption{(Color online) Raw $H(e,e^{\prime}\gamma\gamma)X$ distribution in the [$t_{\rm min}-t$, $Q^2$] plane with cuts for Kin3. The vertical lines delimit the bins we chose in $t_{\rm min}-t$ for our analysis. Superimposed is the ($t_{\rm min}-t$) resolution for each alternate bin, showing that each bin is larger than the resolution.}
    \label{Q2vstmint}
  \end{figure}

  The experimental data used for the analysis have the kinematical coverage shown in Figure \ref{Q2vsxB}. 
  The analysis includes a complete simulation of the resolution and acceptance of the HRS, the external and internal radiative effects on the incident and scattered electron, and a {\sc geant} based simulation of the acceptance and response of the ${\rm PbF}_2$ array. 
  Simulation events are generated uniformly in the target vertex $v$ along the beam line, and uniformly in a phase space $\Delta^5\Phi$.
  This results in well defined values of $\theta_{\pi}^{\rm c.m.}$ in each bin. 
  The $\Delta t$ bins are the same in the generation and experimental phase spaces, but  resolution and radiative effects can cause the migration of events from one bin to one of its neighbors (Figure \ref{Q2vstmint}).
  Rather than extracting average cross sections in the experimental bins, we use the simulation and the theoretical form of Eq. \eqref{dsigmavdt_compact} to directly extract differential cross sections from the experimental yields.
  
  We divide the acceptance into 24 equal bins in $\phi_{\pi} \in [0,2\pi]$ and 8 bins in $t_{\rm min}-t \in [0,0.3] \; {\rm GeV}^2$ for both the helicity dependent and independent parts of the cross section.
  A bin $j_d$ in the kinematic variables reconstructed by the detector is defined by the limits
  $\phi_{\pi} \in[\phi(j_d),\phi(j_d)+\Delta\phi(j_d)]$, 
  $(t_{\rm min}-t)\in [(t_{\rm min}-t)(j_d),(t_{\rm min}-t)(j_d)+\Delta(t_{\rm min}-t)(j_d)]$, {\em etc}.  
  The statistics $\Delta N(j_d)$ in a bin $j_d$ are determined by the physical cross section
  at the vertex convoluted with the detector response:
  \begin{equation}
    \begin{aligned}
      & \Delta N(j_d) 
      \nonumber \\
      &= {\cal L}u \int_{\Delta x_d} dx_d \int_{\Delta x_v} dx_v {\cal R}(x_d,x_v)   \sum_{\Lambda} {\cal F}_{\Lambda}(x_v)r_{\Lambda}
    \end{aligned}
  \end{equation}
  where $x_v$ summarizes the reaction vertex variables, $x_d$ summarizes the reaction vertex variables as reconstructed in the detector, $\Delta x_d$ summarizes the range of integration for bin $j_d$, $\Delta x_v$ summarizes the range of integration for all bins $j_v$, ${\cal L}u$ is the integrated luminosity, and ${\cal R}(x_d,x_v)$ is the probability distribution for an event originating at the vertex with kinematics $x_v$ to be reconstructed by the detector with vertex kinematics $x_d$. This expresses the effects of detector resolution, internal and external radiation, detector efficiency, and anything else that could migrate events from vertex kinematics $x_v$ to the detector kinematics $x_d$.
  For the analysis and simulation, the integral is split into a sum over the bins $\Delta x_v$
  in the kinematic variables at the reaction vertex:
  \begin{equation}
    \begin{aligned}
      &\Delta N(j_d)
      \nonumber \\
      & = {\cal L}u  \int_{\Delta x_d} dx_d \sum_{j_v} \int_{\Delta x_{v \in {\rm bin} \; j_v}} dx_v {\cal R}(x_d,x_v)   \sum_{\Lambda} {\cal F}_{\Lambda}(x_v)r_{\Lambda}
    \end{aligned}
  \end{equation}
  Because the functions ${\cal F}_{\Lambda}(x_v)$ contain the main part of the dependence on the variables at the vertex, the quantity $r_{\Lambda}$ in a bin $\Delta x_v$ will be assimilated to its average $\langle r_{\Lambda} \rangle_{x_v} \equiv r_{v,\Lambda}$ in this bin.
  Then, the last equation can be summarized in a vector notation:
  \begin{equation}
    \Delta N(j_d) = \sum_{j_v} K_{j_d,j_v}^{\Lambda} r_{j_v,\Lambda},
  \end{equation}
  with
  \begin{equation}
    K_{j_d,j_v}^{\Lambda} = 
    {\cal L}u  \int_{\Delta x_d} \int_{\Delta x_{v \in {\rm bin} \; j_v}} {\cal R}(x_d,x_v) {\cal F}_{\Lambda}(x_v)dx_d dx_v\,.
  \end{equation}
  We then replace the integration by a summation over the simulated events $i$:
  \begin{equation}
    K_{j_d,j_v}^{\Lambda} = {\cal L}u  
    \sum_{i\in\{j_v,j_d\} }
    \frac{{\cal F}_{\Lambda}(x_v)}{N_{\rm gen}} \Delta^5 \Phi,
  \end{equation}
  where the sum is over events originating in vertex bin $j_v$ and reconstructed in bin $j_d$.
  $N_{\rm gen}$ is the number of events generated in the simulation and $\Delta^5 \Phi$ is the total phase-space factor.
  The matrices $K_{j_d,j_v}^{\Lambda}$ are constructed from simulation events, summed over all events within cuts.
  We define $N_d = N^+ + N^-$ with $N^+$ ($N^-$) the number of counts within cuts with positive (negative) electron helicity.
  The cuts are the same for simulation and data (Table \ref{Cuts}).
  The cuts and the corrections are summarized in Tables \ref{Cuts} and \ref{Corrections}, respectively.
  \begin{table}
    \begin{center}
      \caption[]{Cuts applied in the primary extraction. $r$ is the value of the so-called $r$ function. The $r$ function defines the distance of the particle from the acceptance bound, and is positive (negative) if the particle is in (out of) the acceptance \cite{rfunc}. The $M_X^2$ and $E_{\rm thr}$ optimizations are presented in Table \ref{MX2_Ethr_Syst}.}
      \label{Cuts}
      \begin{tabular}{|c|}
	\hline      
	\hline      
	\multicolumn{1}{|c|}{Spectrometer cuts}\\
	$-$6.0 cm $<v<$ +7.5 cm \\
	$|x_{\rm HRS \; plane}|<$ 3.5 cm\\
	(Horizontal collimator) \\
	$|y_{\rm HRS \; plane}|<$ 7.0 cm\\
	(Vertical collimator) \\
	$|k^{\prime}-p_{\rm HRS}|/p_{\rm HRS} < 4.5\% $ \\
	$r>$ +0.005 m\\
	\hline
	\multicolumn{1}{|l|}{Calorimeter Cuts}\\
	$-$15.0 cm $<x_{\rm calo}<$ +12.0 cm \\
	$|y_{\rm calo}|<$ 15.0 cm \\
	\hline
	\multicolumn{1}{|l|}{Physics Cuts}\\
	105 MeV $<m_{\gamma \gamma}<$ 165 MeV \\
	\hline
	\hline
      \end{tabular}
    \end{center}
  \end{table}

  A $\chi^2$ is built, assuming that the statistical error on the simulation is much smaller than the statistical error of the data:
  \begin{equation}
    \chi^2 = \sum_{j_d} \frac{\left(N_d - \sum_{j_v}K_{j_d,j_v}^{\Lambda}
    r_{j_v,\Lambda}\right)^2}{N_d} .
  \end{equation}
  The minimization of $\chi^2$ with respect to the unknown quantities $r_{j_v,\Lambda}$ results in  a linear system from which the $r_{j_v,\Lambda}$ are extracted.
  To be fully consistent, one of the two quantities in the numerator has to be corrected for some instrumental systematic effects (Table \ref{Corrections}).
  Note that all vertex bins populate experimental bins, but the detector bin at the largest experimental bin in $(t_{\rm min} - t)$ can receive  contributions from larger values of
  $(t_{\rm min} - t)$, not generated in the simulation.
  Hence, although we extract an  $r_{j_v,\Lambda}$ value for the last bin, we do not include it in our results,
  its role is only to populate the lower $(t_{\rm min} - t)$ bins.

  \begin{table}[htbp]
    \begin{center}
      \caption[]{Correction factors applied in the data analysis.  The radiative correction factor
      is the combination of the virtual radiative correction factors (vertex renormalization and
      vacuum polarization) and the cut-off independent real radiation effects (Sec. \ref{sec6}).
      }
      \label{Corrections}
      \begin{tabular}{|l c c|}
	\hline
	\hline
	Correction & Kin3 & Kin2\\
	\hline
	Multitracks in HRS & 1.079 & 1.099 \\
	Triple cluster in calorimeter & 1.035 & 1.020 \\
	Radiative correction & 0.91 $\pm$ 0.02 & 0.91 $\pm$ 0.02 \\
	\hline
	\hline
      \end{tabular}
    \end{center}
  \end{table}

  The average values of the kinematic variables $Q^2$, $\epsilon$, $x_{\rm Bj}$, $W$, $t$, $t_{\rm min}$, {\it etc.}, in a bin at the vertex are
  \begin{equation}
    \overline{x}_{j_v} = \frac{\sum_{i \in \Delta x_v} x_v K_{j_d,j_v}^{\Lambda}r_{j_v,\Lambda}}{\sum_{i \in \Delta x_v} K_{j_d,j_v}^{\Lambda}r_{j_v,\Lambda}}.
  \end{equation}
  Because the $r_{j_v,\Lambda}$ are by construction constant over the bin $\Delta x_v$ and the integrals of ${\cal F}_{TL}$, ${\cal F}_{TT}$, and ${\cal F}_{TL^{\prime}}$ cancel when integrating over $\phi_{\pi}$, we can write
  \begin{equation}
    \overline{x}_{j_v} = \frac{\sum_{i \in \Delta x_v} x_v K_{j_d,j_v}^{T+ \epsilon_L L}}{\sum_{i \in \Delta x_v} K_{j_d,j_v}^{T+ \epsilon_L L}}.
  \end{equation}
  These values are summarized in Table \ref{meanvars} for quantities independent of the $(t_{\rm min}-t)$ bin and in Table \ref{meanvarsbintmint} for quantities depending on the $(t_{\rm min}-t)$ bin.
  
  Finally, the cross sections at the point $ \overline{x}_{j_v}$ in a bin $j_v$ are obtained by
  \begin{equation}
    \frac{d\sigma_{\Lambda}}{dt} = {\cal F}_{\Lambda} (\overline{x}_{j_v}) r_{j_v,\Lambda}.
  \end{equation}
  The results are displayed in Tables \ref{XsecTab_stat_systErrs_Q2dep} and \ref{XsecTab_stat_systErrs_xBdep}. 
  The first table shows the results for the two kinematics Kin2 and Kin3, which cover the full kinematic range of the experiment, resulting in two domains of different $Q^2$, at constant $x_{\rm Bj}$.
  The second table shows the results for the two kinematics Kin$X$2 and Kin$X$3, which only cover the domain between the two horizontal lines in Figure \ref{Q2vsxB}, in order to have two domains of different $x_{\rm Bj}$ at constant $Q^2$.
  
  \section{radiative corrections}
  \label{sec6}
  
  The external radiative effects on the incident electron, and internal real radiative effects at the vertex
  are treated in the equivalent radiator approximation \cite{RadCorr1, RadCorr2}.
  Preradiation is modeled by generating an event-by-event energy loss $\Delta E_{\rm in}$ of the incident electron ($E_0$) following a distribution ($b \simeq 4/3$):
  \begin{equation}
    I_{\rm in}(E_0, \Delta E_{\rm in}, t_{\rm in}) = \frac{bt_{\rm in} + \delta_S/2}{\Delta E_{\rm in}} \left[ \frac{\Delta E_{\rm in}}{E_0} \right]^{bt_{\rm in} + \delta_S/2}
  \end{equation}
  with
  \begin{equation}
    \delta_S = \frac{2\alpha}{\pi} \left[ \ln\frac{Q^2}{m_e} - 1 \right]
  \end{equation}
  where $t_{\rm in}$ is the event-by-event target thickness (in radiation lengths) traversed by the electron
  before the scattering vertex. 
  The Schwinger term $\delta_S$ models the internal pre-radiation.
  The scattered energy at the vertex is $E_v^{\prime} = E_0-\Delta E_{\rm in}-Q^2/(2 M_p x_{\rm Bj})$.
  Internal post-radiation is modeled by a similar distribution in the post-radiated energy $\Delta E_{\rm out}$:
  \begin{equation}
    I_{\rm out} = \frac{ \delta_S/2}{\Delta E_{\rm out}} \left[ \frac{\Delta E_{\rm out}}{E_v^{\prime}} \right]^{\delta_S/2}
  \end{equation}
  These radiative effects are treated within the peaking approximation.
  External post-radiation by the scattering electron is modeled with the {\sc geant}3 simulation.
  Kinematic shifts ({\em e.g.}, in either the norm and direction of $\vec{q}$) from external and internal radiations are fully included in the simulation and thereby unfolded from the extracted cross sections.
  
  In addition to these radiative effects incorporated into our Monte Carlo, we correct the data for 
  internal virtual radiation (vacuum polarization and vertex renormalization effects) as well as the cut-off independent
  effect of unresolvable soft real radiation. 
  These contributions are calculated by the following terms, respectively \cite{Vanderhaeghen:2000ws}:
  \begin{align}
    \delta_{\rm vacuum} &=  \frac{2\alpha}{ 3 \pi}
    \left[ \ln\left(\frac{Q^2}{ m_e^2}\right) - \frac{5}{ 3} \right]
    \nonumber \\ 
    \delta_{\rm vertex} &= \frac{ \alpha}{\pi}
    \left[ \frac{3}{ 2} \ln\left( \frac{Q^2}{ m_e^2}   \right) - 2 - 
      \frac{1}{ 2}  \ln^2\left( \frac{Q^2}{ m_e^2} \right) 
      + \frac{\pi^2}{ 6} \right]
    \nonumber \\
    \delta_{\rm real,0} &= \frac{ \alpha}{ \pi} \left[
      -{1\over 2} \ln^2\left({E\over E'}\right) 
      + {1\over 2} \ln^2\left({Q^2\over m_e^2}\right) - {\pi^2\over 3}
      \right. \nonumber \\
      &\qquad \left.
      +  {\rm Sp}\left(\cos^2{\frac{\theta_e}{2}}\right) \right] ,
  \end{align}  
  where $\rm{Sp}(\cos^2{\,\theta_e/2})$ is the Spence function.
  After an approximate resummation, the correction we apply to the raw counts (to obtain the equivalent Born approximation cross section) is
  \begin{equation}
    {\rm radcorr} = e^{-\delta_{\rm vertex}-\delta_{\rm real,0}} \left(1-\delta_{\rm vacuum}/2 \right)^2 
  \end{equation}
  The numerical values for our kinematics are tabulated in Table \ref{Corrections}.

  \section{Systematic Errors}
  \label{sec7}

  Two classes of inclusive hadronic electroproduction channels compete with the exclusive $H(e,e^{\prime}\pi^0)p$ reaction: 
 the $H(e,e^{\prime}\pi^0)N\pi, N\pi \pi,...$ channels, with a threshold at $M_X^2 = (M_p + m_{\pi})^2 = 1.15 \; {\rm GeV}^{2}$ and the $H(e,e^{\prime}\pi^0)\gamma p$ channel.
  The first class includes $N^*$ and non-resonant $N\pi$ production in the final state, and diffractive $\rho^+ \rightarrow \pi^+ \pi^0$ production via the $ep \rightarrow e\rho^+ n$ reaction. 
  All these channels can be observed in a missing-mass squared distribution (Figure \ref{MX2distrib}).
  The $H(e,e^{\prime}\pi^0)\gamma p$ channel originates from the diffractive $ep \rightarrow ep \omega$ reaction, with a 8.5\% branching-ratio decay channel \cite{PDG}.
  In our acceptance, the $(e,e^{\prime}\pi^0)$ missing-mass squared threshold for exclusive $\omega$ electroproduction is $1.0 \, {\rm GeV}^{2}$, thus slightly lower than the $N \pi$ threshold of $1.15 \; {\rm GeV}^{2}$.
  However, based on $ep \rightarrow ep \omega$ measurements performed by \cite{omegaHallB}, the expected background of $\omega \pi^0 \gamma$ events for $M_X^2 < 1.15 \; {\rm GeV}^{2}$ is less than 1\% of the exclusive $H(e,e^{\prime}\pi^0) p$ yield in all $t_{\rm min}-t$ bins.
  
\begin{figure}
    \centering
    \includegraphics[width = 0.9\linewidth]{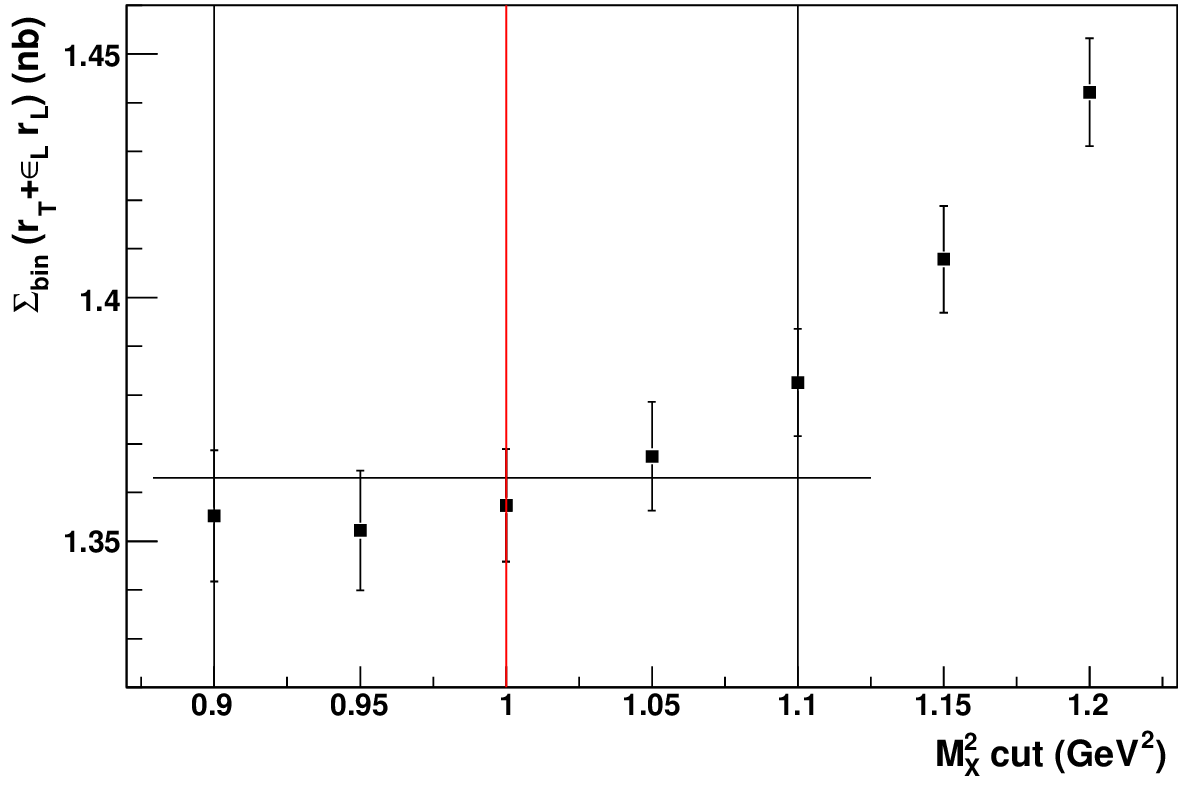}
    \caption{(Color online) Total cross section integrated over $t_{\rm min}-t$ and $\phi_\pi$, for Kin3, as a function of the $M_X^2$ cut. The vertical lines indicate, from left to right, the minimal, optimal, and maximal $M_X^2$ cut values of the stability domain.}
    \label{MX2CutStabilityKin3}
  \end{figure}

\begin{figure}
    \centering
    \includegraphics[width = 0.9\linewidth]{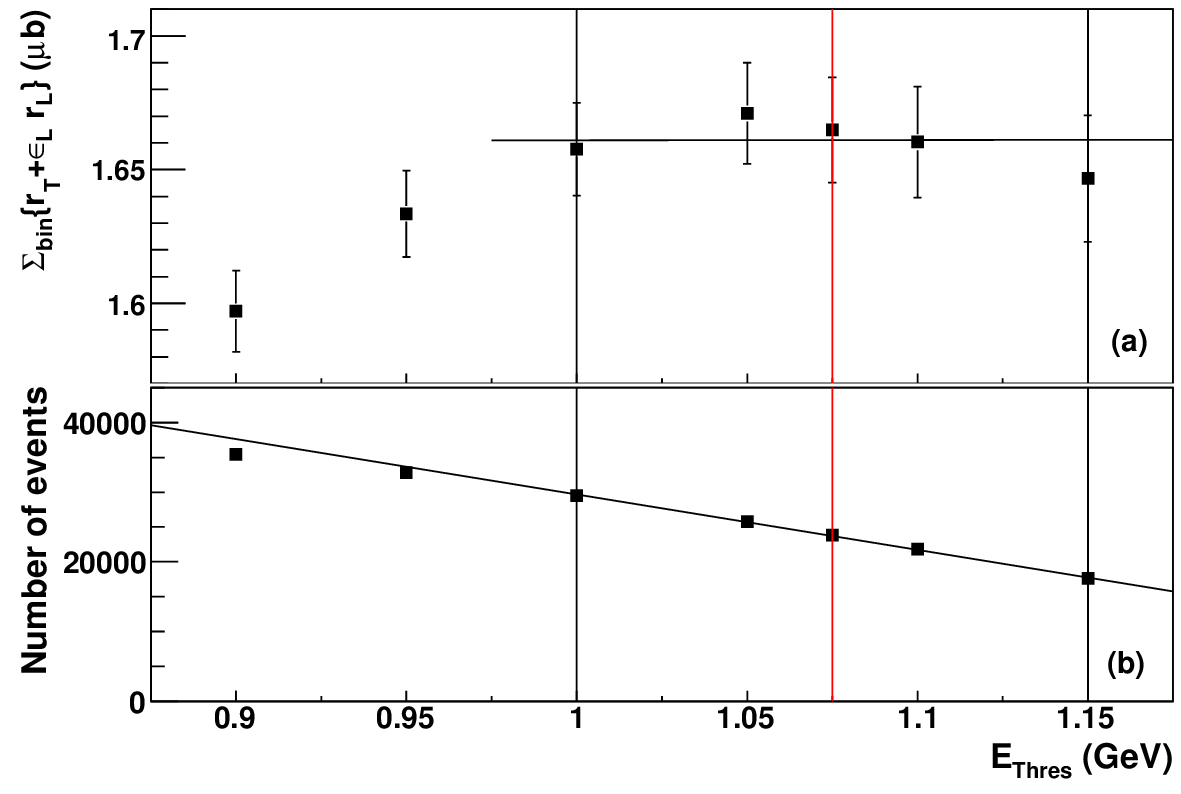}
    \caption{(Color online) (a) Total cross section integrated over $t_{\rm min}-t$ and $\phi_\pi$, for Kin2, as a function of $E_{\rm thr}$. The vertical lines indicate, from left to right, the minimal, optimal, and maximal $E_{\rm thr}$ values of the stability domain (see Table \ref{MX2_Ethr_Syst} for Kin3 values). (b) The number of events as a function of $E_{\rm thr}$. The stability domain for $E_{\rm thr}$ shows the statistics linearly decreasing with $E_{\rm thr}$.}
    \label{EthrStabilityKin2}
  \end{figure}

  The systematic errors in the extraction method are due to the cut on the missing-mass squared $M_X^2$ and on the calorimeter threshold $E_{\rm thr}$.
  The stability of the results is checked by varying each cut in turn.
  The variation in the estimator
  \begin{equation}
    R = \sum_{{\rm bin}=0}^{6} (r_T+\epsilon_L r_L)
  \end{equation}
  is used to quantify the systematic errors.
  %
  \begin{enumerate}
  \item{For the exclusivity ($M_X^2$) cut, we consider the stability interval from $0.9$ to $1.10$ GeV$^2$ in the
  $M_X^2$ cut.  At the high end we expect the cross section to have contributions from inelastic final states (Figure \ref{MX2distrib}).  
  At the low end, we are removing roughly half of the statistics, and we become progressively more sensitive to the 
  experimental line shape.  The stability of the exclusivity cut ({\em e.g.} for Kin3) is plotted in 
  Figures~\ref{MX2CutStabilityKin3}. The cuts and variation are listed in Tables 
  \ref{MX2_Ethr_Syst} and  \ref{Expsyst}.
 In each case, this study is performed with $E_{\rm thr}$ fixed at 1.0 GeV.}
  \item{For the calorimeter threshold $E_{\rm thr}$, the stability of $R$ is expected when the software threshold is fixed above the hardware threshold. Above the hardware threshold, the cut is directly correlated with the $\pi^0 \rightarrow \gamma \gamma$ decay phase space, and the number of events decreases linearly with $E_{\rm thr}$. This comes from the isotropic decay of the pion, leading to a flat energy distribution of each decay photon.  Figure \ref{EthrStabilityKin2} shows for Kin2, the quantity $R$ along with the raw number of counts. The stability is indeed no longer observed when the statistics are not linear with the threshold, meaning the hardware threshold competes with the analysis threshold. The same behavior is shown for Kin3. For both kinematics, the systematic error coming from the calorimeter threshold is evaluated as $\pm1\%$.}
  \end{enumerate}

  The optimal cut is set in the middle of the stability interval (see Figures \ref{MX2CutStabilityKin3} and \ref{EthrStabilityKin2}).
  The stability interval bounds and the optimal values for the $M_X^2$ cut and $E_{\rm thr}$ are listed in Table \ref{MX2_Ethr_Syst} for both kinematics.
  %
  %
  %
  \begin{table}
    \begin{center}
      \caption[]{Values of the $M_X^2$ cut and $E_{\rm thr}$ defining the global cross-section stability domain. Minimum and maximum are the bounds of this domain, and optimum is the cut value set in the middle of the stability interval.}
      \label{MX2_Ethr_Syst}
      \begin{tabular}{|l|c|c|c|}
	\hline
	\hline
	Variable & Minimum & Optimum & Maximum \\
	\hline
	\multicolumn{4}{|c|}{Kin3/Kin$X$3} \\
	\hline
	$M_X^2$ cut (${\rm GeV}^2$) & 0.90 & 1.00 & 1.10\\
	\hline
	$E_{\rm thr}$ (GeV) & 1.20 & 1.275 & 1.35 \\
	\hline
	\multicolumn{4}{|c|}{Kin2/Kin$X$2} \\
	\hline
	$M_X^2$ cut (${\rm GeV}^2$) & 0.90 & 1.00 & 1.10\\
	\hline
	$E_{\rm thr}$ (GeV) & 1.00 & 1.075 & 1.15 \\
	\hline
	\hline
      \end{tabular}
    \end{center}
  \end{table}

  The reduced structure functions $r_{\Lambda}$ are extracted at the optimal value of the cuts.
  For the structure functions implied in $\phi_{\pi}$ dependences, systematic errors are taken as the rms difference between the $r_{\Lambda}$ computed at the optimum cuts and the $r_{\Lambda}$ computed at each of the four extremities of the stability domain.
  
  All instrumental sources of systematic errors are shown along with the analysis systematic errors in Table \ref{Expsyst}.
  \begin{table}[htbp]
    \begin{center}
      \caption[]{Experimental systematic errors. The first ``Total quadratic'' row shows the quadratic sum of all experimental helicity-independent systematic errors. The second ``Total quadratic'' row shows the quadratic sum of all experimental systematic errors including helicity-dependent effects.}
      \label{Expsyst}
      \begin{tabular}{|l|c c|}
	\hline
	\hline
	& Kin3 & Kin2 \\
	& Kin$X$3 (\%) & Kin$X$2 (\%) \\
	\hline
	Exclusivity cut &  1.5 & 3.0 \\
	Calorimeter threshold & \multicolumn{2}{c|}{1.0} \\
	HRS acceptance & \multicolumn{2}{c|}{2.2} \\
	Radiative corrections & \multicolumn{2}{c|}{1.5} \\
	Target length & \multicolumn{2}{c|}{0.5} \\
	Hadronic tensor integration & \multicolumn{2}{c|}{0.3} \\
	Multitracks corrections & \multicolumn{2}{c|}{0.1} \\
	3 clusters corrections & \multicolumn{2}{c|}{0.1} \\
	Luminosity & \multicolumn{2}{c|}{0.1} \\
	Dead time & \multicolumn{2}{c|}{0.1} \\
	Particle identification & \multicolumn{2}{c|}{0.1} \\
	\hline
	Total quadratic & 3.3 & 4.2 \\
	\hline
	Beam polarization & \multicolumn{2}{c|}{2.0} \\
	\hline
	Total quadratic & 3.9 & 4.6 \\
	\hline
	\hline
      \end{tabular}
    \end{center}
  \end{table}
  Since all sources of systematic errors are independent, we added them quadratically. 
  This total systematic error is included in Tables \ref{XsecTab_stat_systErrs_Q2dep} and \ref{XsecTab_stat_systErrs_xBdep}.
  
  \section{Results}
  \label{sec8}
  
  The exclusive $\pi^0$ electroproduction cross section and, in particular, the $\phi_{\pi}$ dependences of its separated components were extracted for Kin2, Kin3, Kin$X$2 and Kin$X$3.
  Our statistics allowed us to achieve, for the $\phi_{\pi}$-independent cross section, a statistical precision of 3\% for Kin2 and Kin3, and of 5\% for Kin$X$2 and Kin$X$3.
  This difference is due to the fact that we could use the full statistics for Kin2 and Kin3, whereas less than half of the statistics were available for Kin$X$2 and Kin$X$3.

\begin{figure}
    \includegraphics[width = \linewidth]{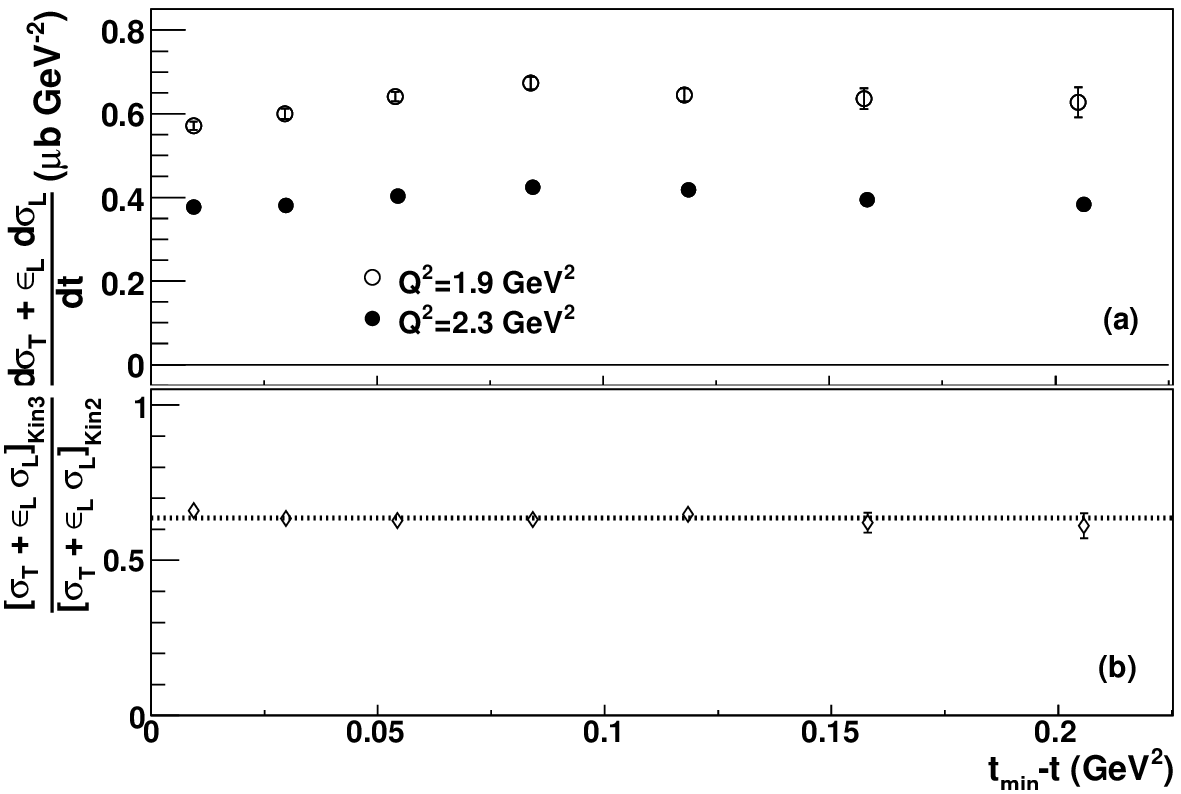}
    \caption{(a) Separated $H(e,e^{\prime}\pi^0)p$ cross section $\sigma_T +\epsilon_L \sigma_L$ as a function of $t_{\rm min}-t$ for $x_{\rm Bj} = 0.36$. Error bars represent statistical errors only. (b) Ratio of $\sigma_T +\epsilon_L \sigma_L$ for the two kinematics as a function of $t_{\rm min}-t$. The fit of this ratio (dashed line) indicates the $Q^2$ dependence of the cross section.}
    \label{TepsL_Q2dep}
  \end{figure}

\begin{figure}
    \includegraphics[width = \linewidth]{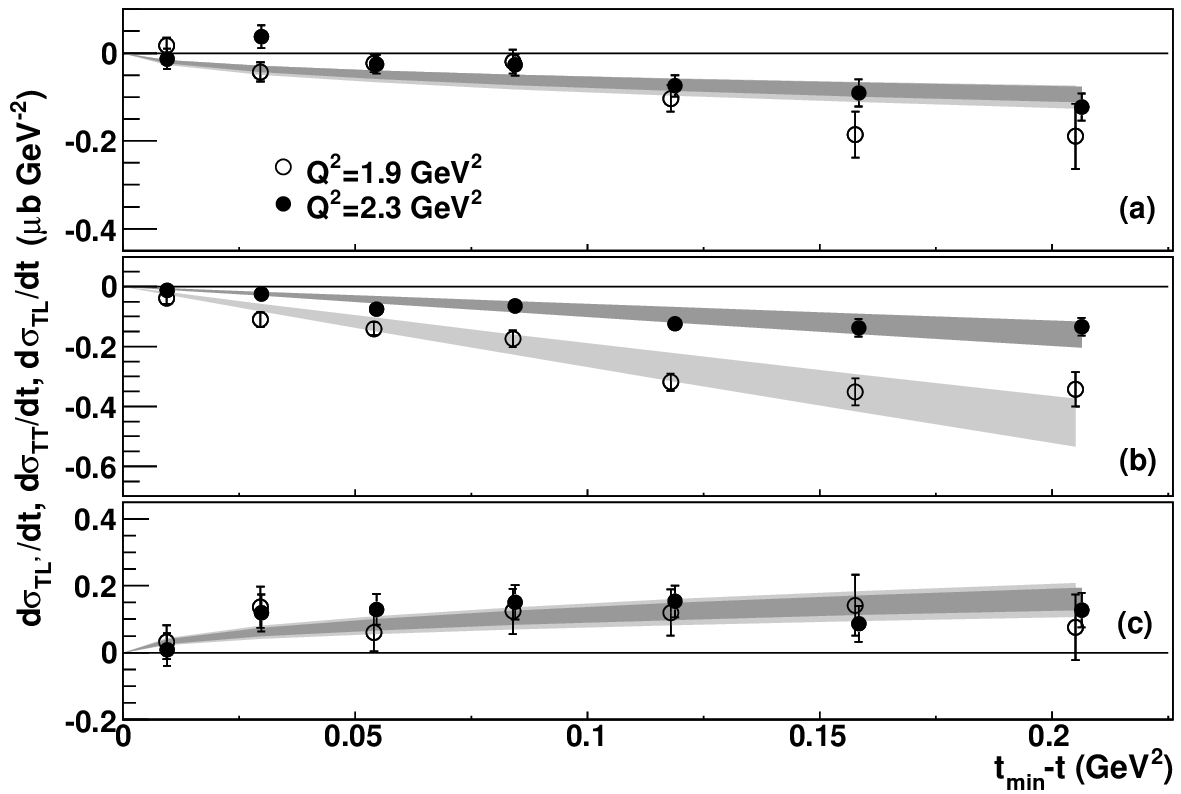}
    \caption{$\sigma_{TL}$ (a), $\sigma_{TT}$ (b), and $\sigma_{TL^{\prime}}$ (c) $H(e,e^{\prime}\pi^0)p$ cross-section components as a function of $t_{\rm min}-t$ for the two $Q^2$-values. Kin2 is represented by the open circles and Kin3 by the solid circles. Error bars represent statistical errors only. The bands (light for Kin2 and dark for Kin3) show fits proportional to $\sin{\,\theta_{\pi}^{\rm c.m.}}$, $\sin^2{\,\theta_{\pi}^{\rm c.m.}}$, and $\sin{\,\theta_{\pi}^{\rm c.m.}}$, respectively. 
    Refer to Table \ref{XsecTab_stat_systErrs_Q2dep} for more detailed cross-section values, with statistical and systematic errors.}
    \label{TL_TT_TLprim_Q2dep}
  \end{figure}

\begin{figure}
    \centering
    \includegraphics[width = \linewidth]{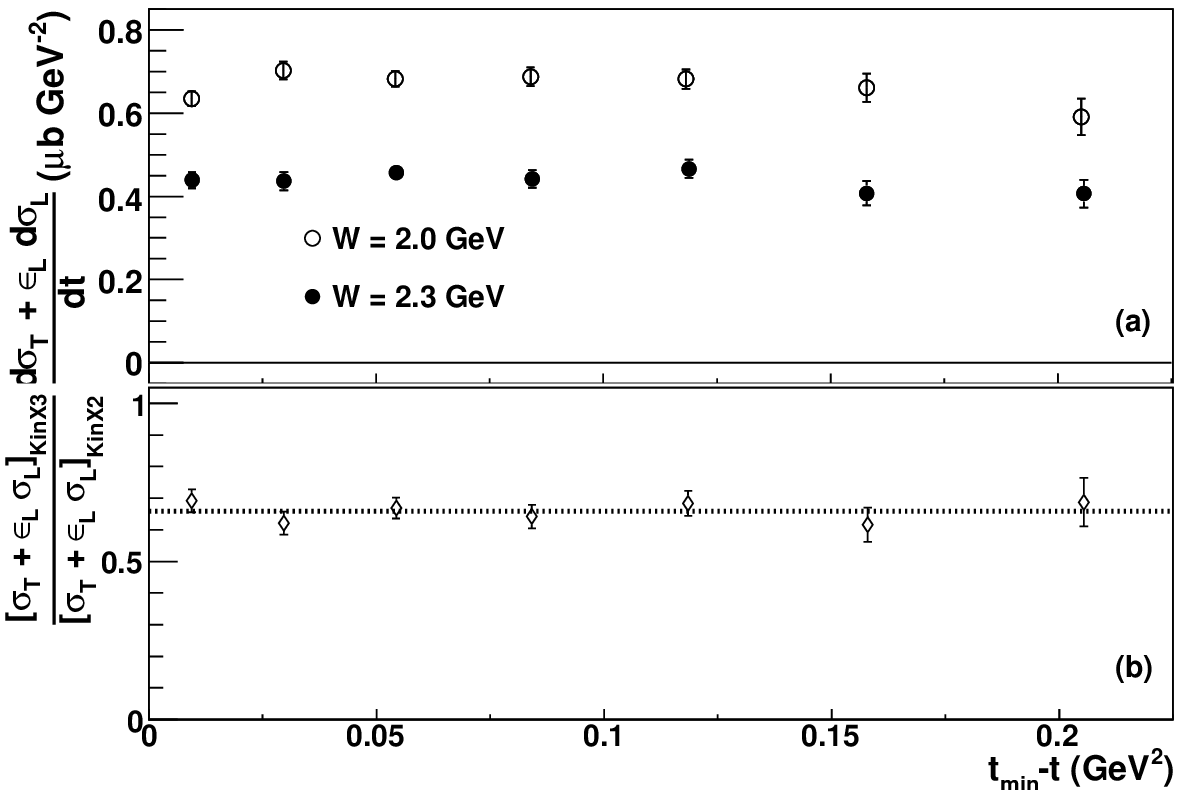}
    \caption{(a) Separated $H(e,e^{\prime}\pi^0)p$ cross section $\sigma_T +\epsilon_L \sigma_L$ as a function of $t_{\rm min}-t$ for $Q^2 = 2.1 \; {\rm GeV}^2$. Error bars represent statistical errors only. (b) Ratio of $\sigma_T +\epsilon_L \sigma_L$ for the two kinematics as a function of $t_{\rm min}-t$. The fit of this ratio (dashed line) indicates the $W$ dependence of the cross section.}
    \label{TepsL_Wdep}
  \end{figure}

\begin{figure}
    \includegraphics[width = \linewidth]{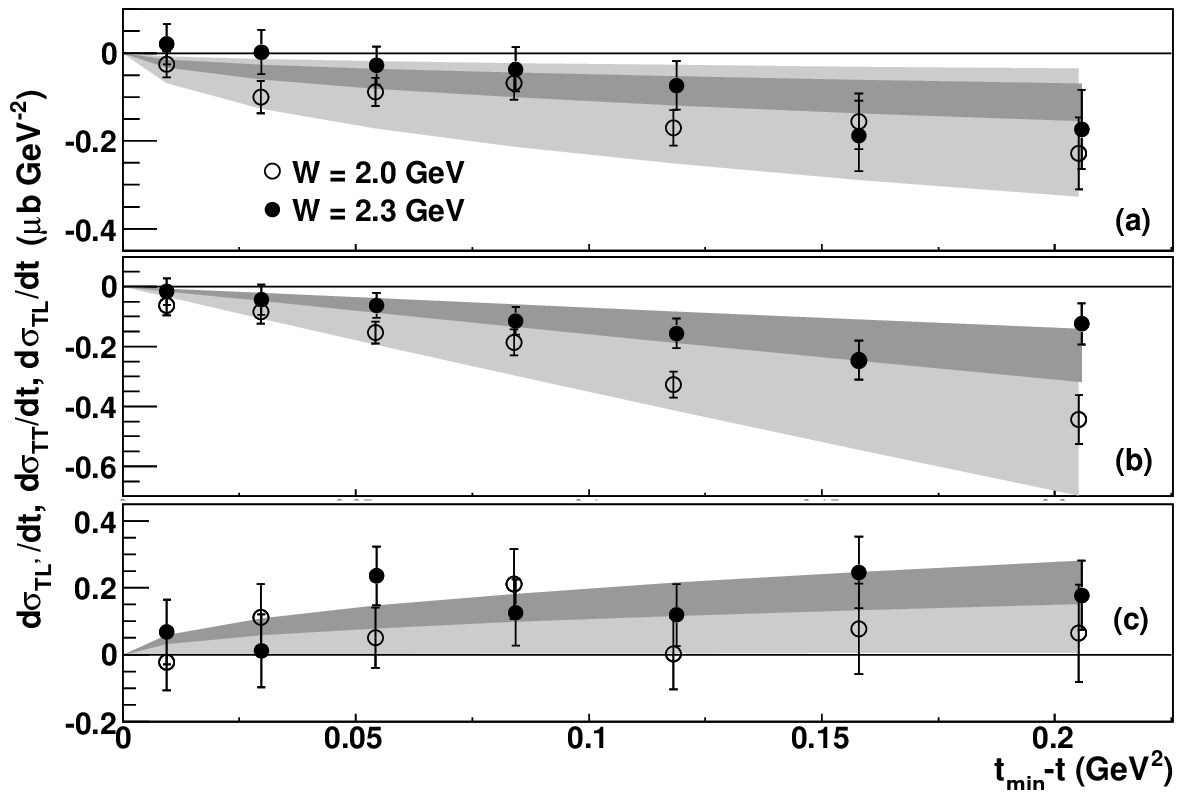}
    \caption{$\sigma_{TL}$ (a), $\sigma_{TT}$ (b), and $\sigma_{TL^{\prime}}$ (c) $H(e,e^{\prime}\pi^0)p$ cross-section components as a function of $t_{\rm min}-t$ for the two $x_{\rm Bj}$-values. Kin$X$2 is represented by the open circles and Kin$X$3 by the solid circles. Error bars represent statistical errors only. The bands (light for Kin$X$2 and dark for Kin$X$3) show fits proportional to $\sin{\,\theta_{\pi}^{\rm c.m.}}$, $\sin^2{\,\theta_{\pi}^{\rm c.m.}}$, and $\sin{\,\theta_{\pi}^{\rm c.m.}}$, respectively. Refer to Table \ref{XsecTab_stat_systErrs_xBdep} for more detailed cross-section values, with statistical and systematic errors.}
    \label{TL_TT_TLprim_Wdep}
  \end{figure}

  Figure \ref{TepsL_Q2dep} shows $\sigma_T + \epsilon_L \sigma_L$ and
  Figure \ref{TL_TT_TLprim_Q2dep} shows $\sigma_{TL}$, $\sigma_{TT}$, and $\sigma_{TL^{\prime}}$ plotted as a function of $t_{\rm min}-t$, both for Kin2 and Kin3.   
  Figure \ref{TepsL_Wdep} shows $\sigma_T + \epsilon_L \sigma_L$ and Figure \ref{TL_TT_TLprim_Wdep} shows $\sigma_{TL}$, $\sigma_{TT}$, and $\sigma_{TL^{\prime}}$, both for Kin$X$2 and Kin$X$3.
  %
  %
  
  We performed fits proportional to $\sin{\,\theta_{\pi}^{\rm c.m.}}$ for $\sigma_{TL}$ and $\sigma_{TL^{\prime}}$, and proportional to $\sin^2{\theta_{\pi}^{\rm c.m.}}$ for $\sigma_{TT}$.
  These fits, including statistical and systematic errors, are shown as bands in Figures \ref{TL_TT_TLprim_Q2dep} and \ref{TL_TT_TLprim_Wdep}, and in Tables \ref{Had_Tens_Q2dep} and \ref{Had_Tens_xBdep}.
  Their reduced $\chi^2$ are below 1.05 for the $Q^2$-dependent data, and below 0.75 for the $x_{\rm Bj}$-dependent data.
  This confirms that the main $t$ dependence of $\sigma_{TL, TL^{\prime}}$, and $\sigma_{TT}$ is given by $\sin{\,\theta_{\pi}^{\rm c.m.}}$ and $\sin^2{\,\theta_{\pi}^{\rm c.m.}}$, respectively.
  
  The lower panel of Figure \ref{TepsL_Q2dep} (respectively, Figure \ref{TepsL_Wdep}) also shows the $Q^2$ dependence (respectively, $x_{\rm Bj}$ dependence) for the total cross section $\sigma_T + \epsilon_L \sigma_L$.      
  To investigate a $Q^2$ or a $x_{\rm Bj}$ dependence, the ratio of $\sigma_T + \epsilon_L \sigma_L$ for the two kinematics is plotted as a function of $t_{\rm min} - t$.
  This ratio is found to be independent of $t$, thus the value of this ratio is fitted by a constant at the $x_{\rm Bj}$- and $Q^2$- values for the two kinematics.

  The dependence of $\sigma_T + \epsilon_L \sigma_L$ in Figures \ref{TepsL_Q2dep} and \ref{TepsL_Wdep} yields the following conclusions:
  \begin{enumerate}
  \item{The ratio $[\sigma_T + \epsilon_L \sigma_L]_{\rm Kin3}/[\sigma_T + \epsilon_L \sigma_L]_{\rm Kin2}$ is flat in $t_{\rm min}-t$ with a reduced $\chi^2$ of 0.33. The ratio is found to be $0.633 \pm 0.009$, indicating a $Q^2$ dependence of the total cross section of about $1/Q^{4.5}$.}
  \item{The ratio $[\sigma_T + \epsilon_L \sigma_L]_{{\rm Kin}X3}/[\sigma_T + \epsilon_L \sigma_L]_{{\rm Kin}X2}$ is also flat in $t_{\rm min}-t$ with a reduced $\chi^2$ of 0.56. This ratio is found to be $0.660 \pm 0.015$, indicating a $W$ dependence of the total cross section of about $1/W^{3.5}$.}
  \end{enumerate}
  The $Q^2$ and $W$ dependences of the relevant quantities [$\sigma_T+ \epsilon_L \sigma_L$, $\sigma_T$, and $\sigma_L$, with our conventions ({\it i.e} Drechsel-Tiator) and VGG conventions] have been summarized in Table \ref{XsecHadTens_Q2Wdeps}.
  \begin{table}[htbp]
    \begin{center}
      \begin{tabular}{|l c c|}
	\hline
	\hline
	Quantity & $Q^2$ dependence & $W$ dependence \\
	\hline
	$\sigma_T + \epsilon_L \sigma_L$ & $(Q^2)^{-2.39 \pm 0.08}$ & $(W)^{-3.48 \pm 0.11}$ \\
	$\sigma_L$ (Drechsel-Tiator) & $(Q^2)^{-0.50 \pm 0.13}$ & $(W)^{-0.46 \pm 0.57}$ \\
	$\sigma_L$ (VGG) & $(Q^2)^{-1.50 \pm 0.08}$ & $(W)^{1.28 \pm 2.52}$ \\
	\hline
	\hline
      \end{tabular}
      \caption[]{$Q^2$ and $W$ dependences for the total cross section and the longitudional cross section with Drechsel-Tiator conventions and with VGG conventions. For $\sigma_L$, the dependences have been evaluated neglecting $\sigma_T$. The $Q^2$ and $W$ dependences of $\sigma_T$ alone ({\it i.e.} assuming $\sigma_L = 0$) are the same as the $Q^2$ and $W$ dependences of $\sigma_T + \epsilon_L \sigma_L$.}
      \label{XsecHadTens_Q2Wdeps}
    \end{center}
  \end{table}

\begin{figure*}
    \begin{center}
      \centering
      \includegraphics[width = 0.8\linewidth]{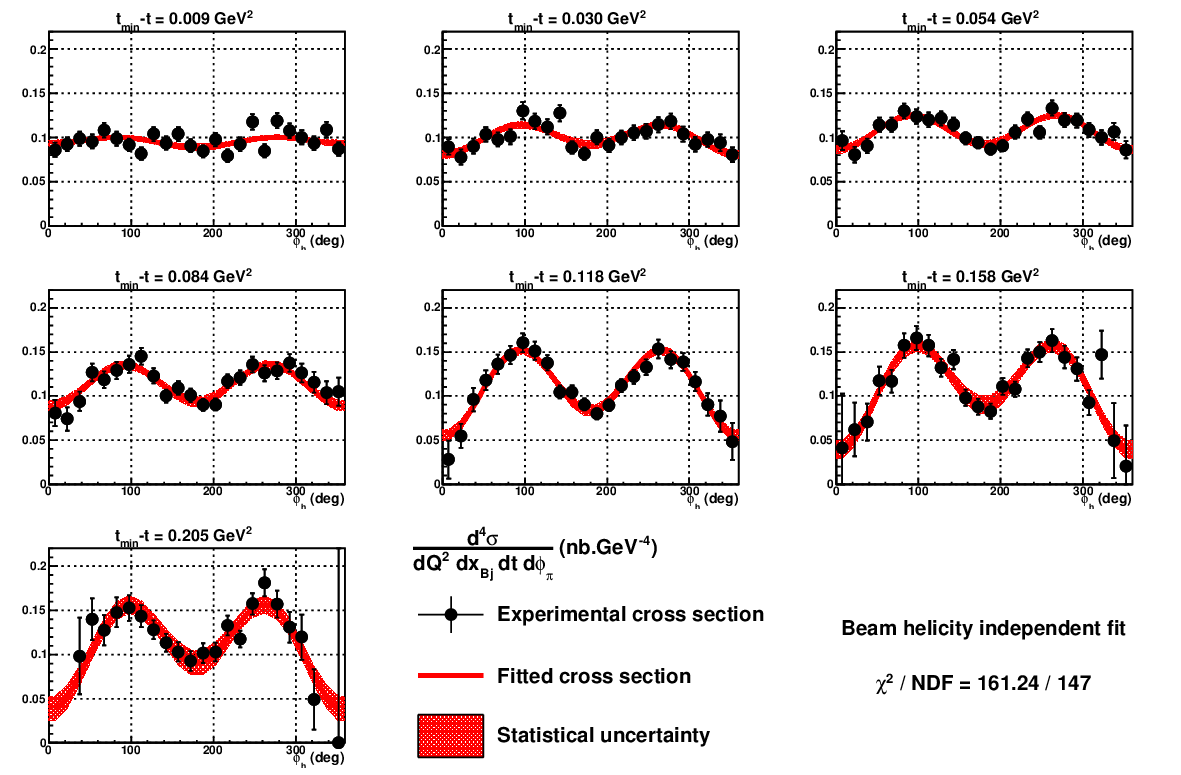}
      \caption{Experimental cross section $d^4 \sigma / dQ^2 dx_{Bj} dt d\phi_{\pi}$ as provided in Eq. \eqref{d4sigExp}, as a function of $\phi_{\pi}$ for each bin in $t_{\rm min}-t$, $\phi_{\pi}$, for $Q^2 = 1.9 \, {\rm GeV}^2$ (black solid points). The red solid curves are $d^4 \sigma_{fit} / dQ^2 dx_{Bj} dt d\phi_{\pi}$ as provided in Eq. \eqref{d4sigFit}. The numerical values are provided in Table \ref{ExpXsecTab_Kin2}.}
      \label{ExpXsecKin2}
    \end{center}
  \end{figure*}

\begin{figure*}
    \begin{center}
      \centering
      \includegraphics[width = 0.8\linewidth]{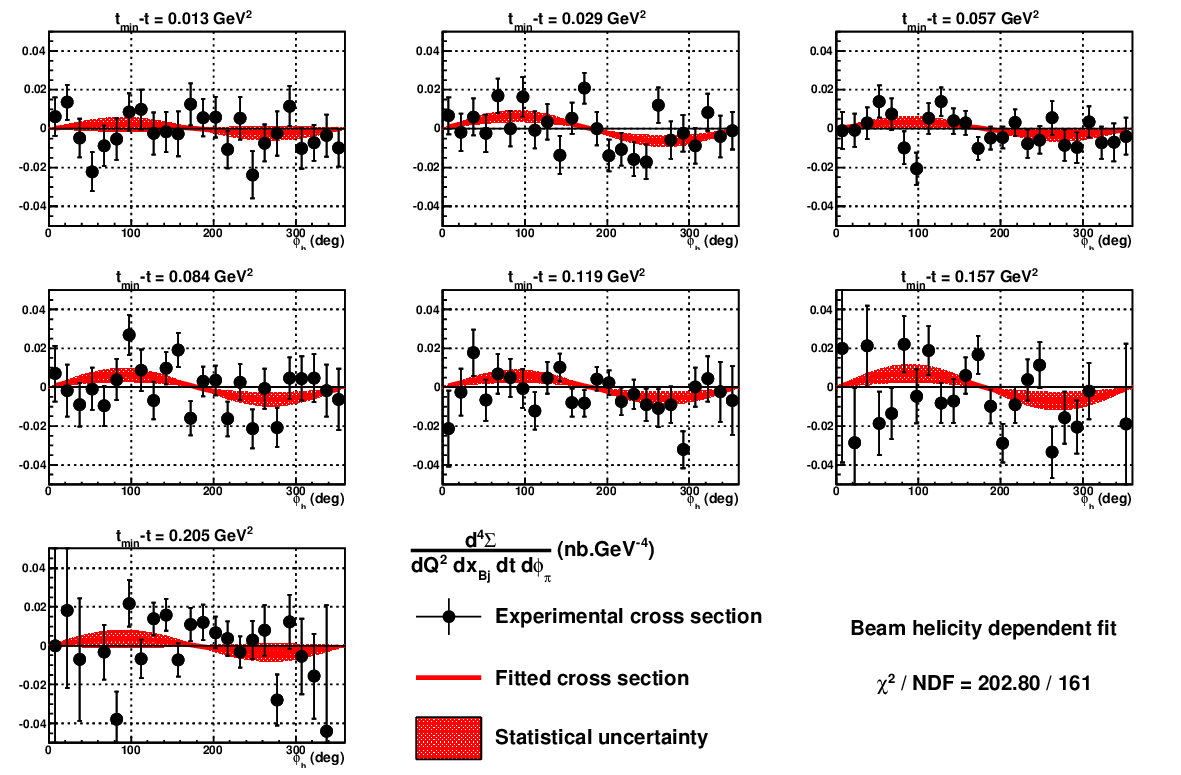}
      \caption{Experimental cross section $d^4 \Sigma / dQ^2 dx_{Bj} dt d\phi_{\pi}$ as provided in Eq. \eqref{d4sigExp}, as a function of $\phi_{\pi}$ for each bin in $t_{\rm min}-t$, $\phi_{\pi}$, for $Q^2 = 1.9 \, {\rm GeV}^2$ (black solid points). The red solid curves are $d^4 \sigma_{fit} / dQ^2 dx_{Bj} dt d\phi_{\pi}$ as provided in Eq. \eqref{d4SigFit}. The numerical values are provided in Table \ref{ExpXsechdepTab_Kin2}.}
      \label{ExpXsechdepKin2}
    \end{center}
  \end{figure*}

\begin{figure*}
    \begin{center}
      \centering
      \includegraphics[width = 0.8\linewidth]{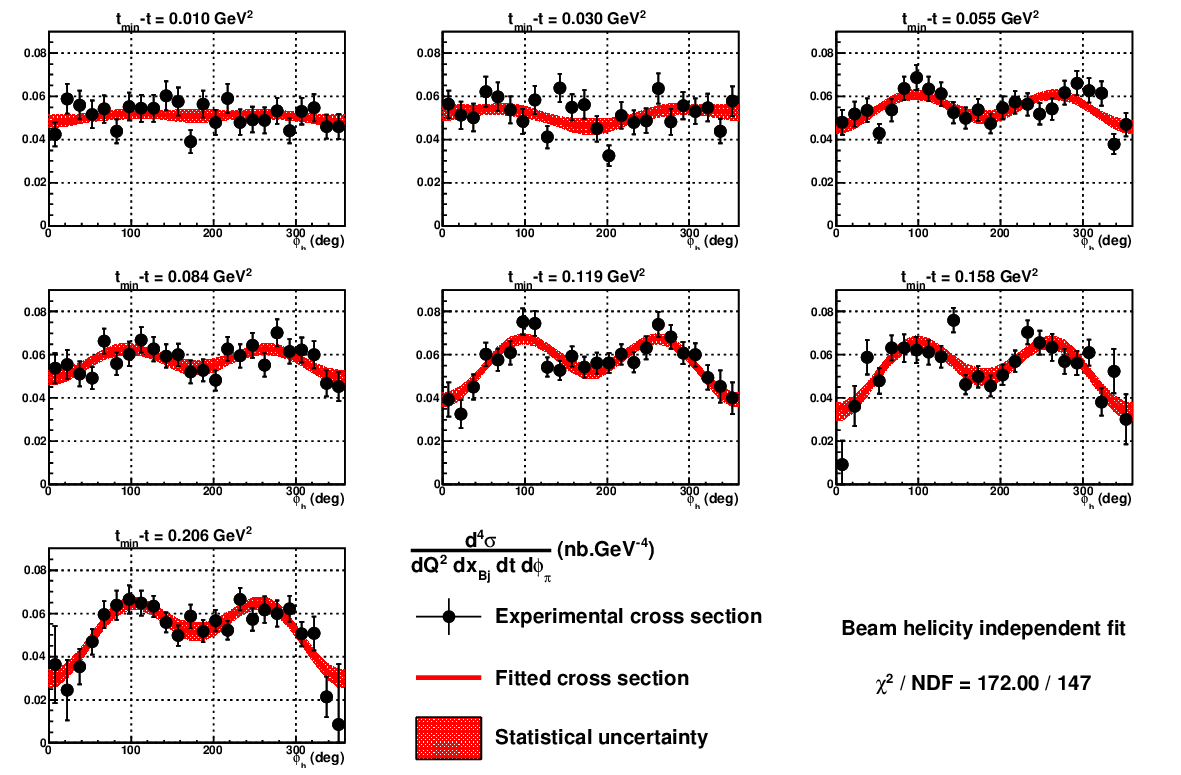}
      \caption{Experimental cross section $d^4 \sigma / dQ^2 dx_{Bj} dt d\phi_{\pi}$ as provided in Eq. \eqref{d4sigExp}, as a function of $\phi_{\pi}$ for each bin in $t_{\rm min}-t$, $\phi_{\pi}$, for $Q^2 = 2.3 \, {\rm GeV}^2$ (black solid points). The red solid curves are $d^4 \sigma_{fit} / dQ^2 dx_{Bj} dt d\phi_{\pi}$ as provided in Eq. \eqref{d4sigFit}. Numerical values are provided in Table \ref{ExpXsecTab_Kin3}.}
      \label{ExpXsecKin3}
    \end{center}
  \end{figure*}

\begin{figure*}
    \begin{center}
      \centering
      \includegraphics[width = 0.8\linewidth]{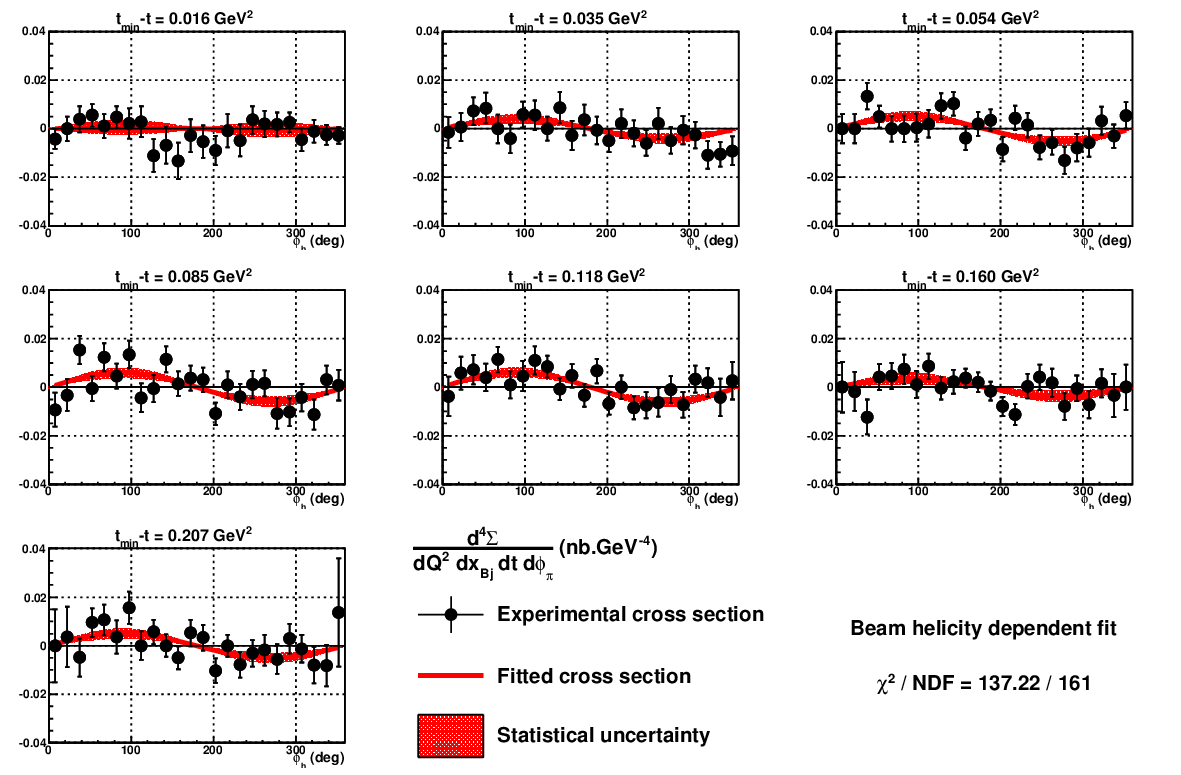}
      \caption{Experimental cross section $d^4 \Sigma / dQ^2 dx_{Bj} dt d\phi_{\pi}$ as provided in Eq. \eqref{d4sigExp}, as a function of $\phi_{\pi}$ for each bin in $t_{\rm min}-t$, $\phi_{\pi}$, for $Q^2 = 2.3 \, {\rm GeV}^2$ (black solid points). The red solid curves are $d^4 \sigma_{fit} / dQ^2 dx_{Bj} dt d\phi_{\pi}$ as provided in Eq. \eqref{d4SigFit}. The numerical values are provided in Table \ref{ExpXsechdepTab_Kin3}.}
      \label{ExpXsechdepKin3}
    \end{center}
  \end{figure*}

  We extract the experimental cross sections 
  \begin{equation}
    \frac{d^4 \sigma}{dQ^2 dx_{Bj} dt d\phi_{\pi}}, \; \frac{d^4 \Sigma}{dQ^2 dx_{Bj} dt d\phi_{\pi}},
    \label{d4sig}
  \end{equation}
  (respectively beam helicity independent and beam helicity dependent) for each bin in $t_{\rm min}-t$ and $\phi_{\pi}$. 
  They are defined, for each vertex kinematic bin $j_v$ in terms of the yield in the corresponding bin $j_d$ as:
  \begin{equation}
    \frac{d^4 \sigma,\Sigma(j)}{dQ^2 dx_{Bj} dt d\phi_{\pi}} = 2\pi \frac{d^5 \sigma,\Sigma(j_v)^{fit}}{dQ^2 dx_{Bj} d\phi_e dt d\phi_{\pi}} \cdot 
    \frac{N(j_d)}{\Delta N (j_d)}.
    \label{d4sigExp}
  \end{equation}
  The experimental counts $N(j_d)$ and simulation counts $\Delta N (j_d)$ are defined previously in the text.
  The five-fold differential cross section are defined as
  \begin{equation}
    \frac{d^5 \sigma_{fit}}{dQ^2 dx_{Bj} d\phi_e dt d\phi_{\pi}} = \sum_{\Lambda \in \{ T+\epsilon_L L, TL, TT\}} {\cal F}_{\Lambda} (\overline{x}_{j_v}) r_{j_v,\Lambda} 
    \label{d4sigFit}
  \end{equation}
  and 
  \begin{equation}
    \frac{d^5 \Sigma_{fit}}{dQ^2 dx_{Bj} d\phi_e dt d\phi_{\pi}} = {\cal F}_{TL^{\prime}} (\overline{x}_{j_v}) r_{j_v,TL^{\prime}}.
    \label{d4SigFit}
  \end{equation}
  The experimental cross sections $d^4 \sigma / dQ^2 dx_{Bj} dt d\phi_{\pi}$ and $d^4 \Sigma / dQ^2 dx_{Bj} dt d\phi_{\pi}$ are plotted in Figs. \ref{ExpXsecKin2} and \ref{ExpXsechdepKin2} and tabulated in Tables \ref{ExpXsecTab_Kin2} and \ref{ExpXsechdepTab_Kin2}, respectively, for Kin2 ($Q^2 = 1.9 \, {\rm GeV}^2$). 
  Corresponding  plots and tables are presented in Figs. \ref{ExpXsecKin3} and \ref{ExpXsechdepKin3} respectively, and Tables \ref{ExpXsecTab_Kin3} and \ref{ExpXsechdepTab_Kin3} for Kin3 ($Q^2 = 2.3 \, {\rm GeV}^2$).

  \section{Discussion}
  \label{sec9}
  
  In the domain in $t_{\rm min}-t$ where we extracted cross sections, the $r_{\Lambda}$ values from Eqs. \eqref{rTL_TLprim} and \eqref{rTT} are constant within statistics, as evidenced by the fits in Figures \ref{TL_TT_TLprim_Q2dep} and \ref{TL_TT_TLprim_Wdep}.
  
\begin{figure}[h!]
    \centering
    \includegraphics[width = 0.9\linewidth]{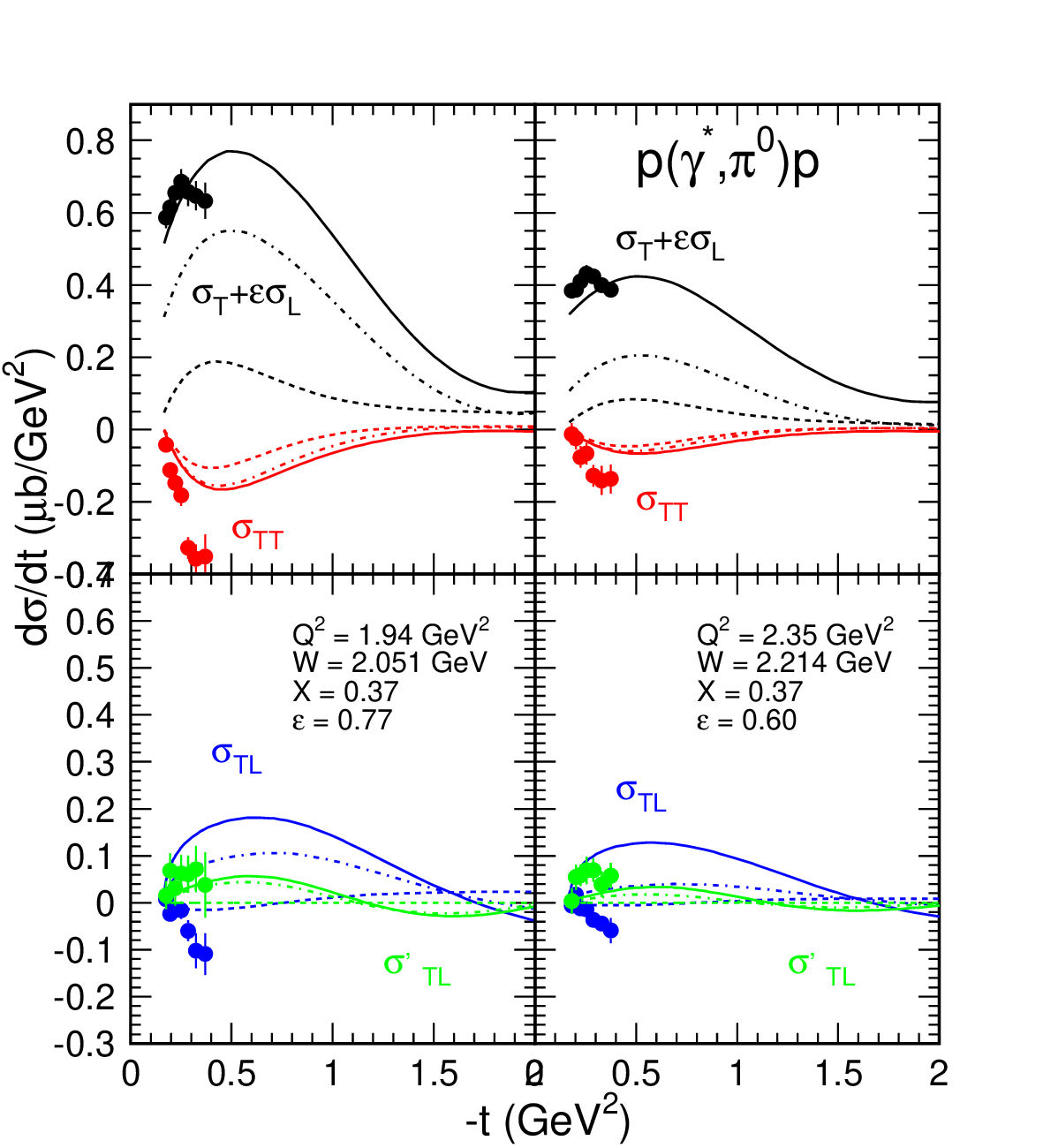}
    \caption{(Color online) New calculations at Kin2 (left panels) and Kin3 (right panels) of the $t$-channel meson-exchange model,
    including charge pion rescattering with $\pi N$ and $\pi\Delta$ intermediate states \cite{LagetNew}. Dashed lines: pole contributions and Pomeron cut alone. Dash-dotted lines: without $\rho \Delta$ cuts. Full lines: $\rho \Delta$ cuts included.}
    \label{LagetFig}
  \end{figure}

  The data we extracted (see the previous section) yield two conclusions with regard to the available models:
  \begin{enumerate}
  \item{ The $t$-channel meson-exchange model of Laget (Figure \ref{LagetFig}) is able to describe $\sigma_T+\epsilon_L \sigma_L$ and $\sigma_{TL^{\prime}}$, but neither $\sigma_{TL}$ nor $\sigma_{TT}$
  \cite{LagetNew}.}
  \item{the $Q^2$ dependence of the cross section (Figure \ref{TepsL_Q2dep} and Table \ref{XsecHadTens_Q2Wdeps}) demonstrates that we are far from the QCD leading twist prediction of $d\sigma_L/dt$, which behaves as $1/Q^6$. On the other hand, it is similar to the $Q^2$ dependence of the transverse cross section for charged pion electroproduction published by Hall C  \cite{HallCpiplus_Long}.}
  \end{enumerate}

  Moreover, the $\pi^0$ has no charge and no spin, so a direct coupling with a virtual photon is suppressed, which removes the pion-pole contribution to the longitudinal cross section.
  This suggests that the transverse $ep\rightarrow~ep\pi^0$ cross section is likely to dominate, and transverse $ep\rightarrow en \pi^+$ cross sections have already been described by quark fragmentation mechanisms usually used to describe semi-inclusive processes.

  T. Horn {\it et al.} measured the exclusive $\pi^+$ electroproduction cross section at $Q^2 = 1.60$ and $2.45 \; {\rm GeV}^2$, with $\sigma_T$ and $\sigma_L$ separation \cite{HallCpiplus_Long}.
  The $t$-channel meson-exchange model by Laget reproduces the $\sigma_L$ component.
  However, the $\sigma_T$ component does not follow the TME model prediction.
  Kaskulov {\it et al.} performed {\sc pythia-jetset} calculations using the Lund model applied to $\pi^+$ transverse cross sections at Hall C kinematics \cite{Lund}. 
  In this model, the virtual photon strikes a quark, with a probability given by the structure functions.
  Due to this, the hadronic system fragments into two jets. 
  The jet engendered by the single quark gives a pion, and the one engendered by the remainder of the nucleon gives the final neutron.
  These calculations applied to Hall C $\pi^+$ transverse cross sections are in excellent agreement with the data.
  This gives evidence that the $\pi^+$ transverse cross section at $Q^2 >1 \; {\rm GeV}^2$ above the resonance region is described by a partonic process.
  This suggests that the present $\pi^0$ data could similarly be described by incoherent scattering on the partonic structure of the nucleon target.

\begin{figure}
    \centering
    \includegraphics[width = 0.9\linewidth]{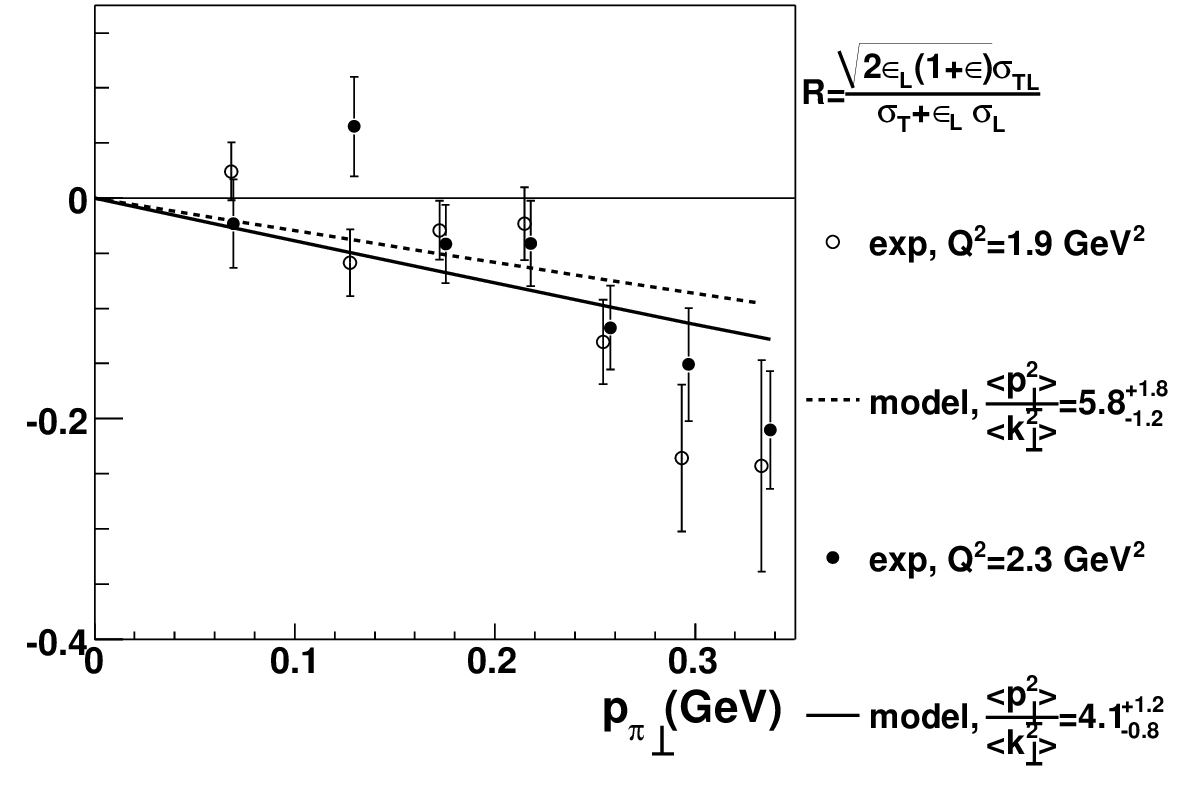}
    \caption{Ratio $\frac{\sqrt{2 \epsilon_L(1+\epsilon)} \sigma_{TL}}{\sigma_T+\epsilon_L \sigma_L}$ for Kin2 (open circles) and Kin3 (solid circles) plotted as a function of $p_{\pi \perp}$. Error bars represent statistical errors only. We fitted to each kinematics a model by Anselmino {\it et al.} in \cite{anselmino} using $\langle p_{\perp}^2 \rangle / \langle k_{\perp}^2 \rangle$ as a free parameter, where $\langle k_{\perp}^2 \rangle$ is the intrinsic transverse momentum of quarks and $\langle p_{\perp}^2 \rangle$ is the transverse momentum transferred during the hadronization process. The reduced $\chi^2$ of the fits are 2.12 for Kin3 and 2.65 for Kin2.}
    \label{TLoverTepsLratio}
  \end{figure}

  For these reasons, we consider our data within the context of semi-inclusive deep inelastic scattering (SIDIS).
  We can try to fit our data with a SIDIS formalism written by Anselmino {\it et al.} \cite{anselmino}.
  Equation (38) of \cite{anselmino} gives the cross section for semi-inclusive production of a pion (valid for any hadron):
  \begin{equation}
    \begin{aligned}
      &\frac{d^5 \sigma^{\ell p \rightarrow \ell \pi X}}{dx_{\rm Bj} dQ^2 dz_{\pi} d^2p_{\pi \perp}} \simeq \sum_{q} \frac{2\pi \alpha^2 e_q^2}{Q^4} f_q(x_{\rm Bj}) D_q^h(z_{\pi})\\
      &\times \left[ 1+(1-y)^2 - 4 \frac{(2-y) \sqrt{1-y} \langle k_{\perp}^2 \rangle z_{\pi} p_{\pi \perp}}{\langle p_{\pi \perp}^2 \rangle \sqrt{Q^2}} \cos \phi_{\pi} \right]\\
      &\times \frac{1}{\pi \langle p_{\pi \perp}^2 \rangle} e^{-p_{\pi \perp}^2/\langle p_{\pi \perp}^2 \rangle}
    \end{aligned}
    \label{eq38anselmino}
  \end{equation}
  %
  where $y = p  q/ p  k$, and $z_{\pi} = p  p_{\pi}/ p  q$ is the fraction of the reaction energy carried by the measured hadron, and the quantities between angle brackets are the standard deviations of transverse momentum distributions, which are approximated as Gaussian.
  $\langle k_{\perp}^2 \rangle$ stands for the parton transverse momentum in the proton, and $\langle p_{\pi \perp}^2 \rangle = \langle p_{\perp}^2 \rangle + z_{\pi}^2 \langle k_{\perp}^2 \rangle$ is the measured transverse momentum of the observed hadron, where $\langle p_{\perp}^2 \rangle$ stands for the transverse momentum of the hadron with respect to the direction of the struck quark.
  The idea is to adjust the ratio of $\cos \,\phi_{\pi}$ over constant term in brackets of Eq. \eqref{eq38anselmino} by adjusting only the parameter $\langle p_{\perp}^2 \rangle/\langle k_{\perp}^2 \rangle$:
  \begin{equation}
     \frac{\sqrt{2\epsilon_L (1+\epsilon)}\sigma_{TL}}{\sigma_T+\epsilon_L \sigma_L} = \frac{4 (2-y) \sqrt{1-y} z_{\pi} p_{\pi \perp}}{\left(\frac{\langle p_{\perp}^2 \rangle}{\langle k_{\perp}^2 \rangle}+z_{\pi}^2 \right) \sqrt{Q^2} [1+(1-y)^2]}
  \end{equation}
  Two conclusions arise from the fits shown in Figure \ref{TLoverTepsLratio}: (1) the minus sign affecting the $\cos \,\phi_{\pi}$ term in the SIDIS model is in agreement with the $\sigma_{TL}$ and (2) $\langle p_{\perp}^2 \rangle$ must be equal to $\sim 5.0 \times \langle k_{\perp}^2 \rangle$ to reproduce the data.

  The authors of \cite{anselmino} adjusted their model to semi-inclusive data. They give $\langle k_{\perp}^2 \rangle = 0.25 \; {\rm GeV}^2$ and $\langle p_{\perp}^2 \rangle = 0.20 \; {\rm GeV}^2$, giving a ratio $\langle p_{\perp}^2 \rangle / \langle k_{\perp}^2 \rangle \sim 0.8$.
  However, they extracted these values in the inclusive region, implying a high multiplicity of particles, whereas in our data, the multiplicity of particles is unity.
  Typically, Anselmino {\it et al.} fit their model with data covering the range $0.1 < z_{h} <1.0$, with most of the statistics within $z_h < 0.4$, whereas our data are within $z_h>0.9$.
  Furthermore, Kaskulov {\it et al.} \cite{Lund} used a value of 1.4 $\rm{GeV}^2$ for the rms transverse momentum of partons in their fit of the Hall C $\pi^+$ data.
  The exclusive limit of SIDIS could be defined by a SIDIS-inspired model applicable to data at $z_{\pi} \rightarrow 1.0$ or, more practically, when the measured hadron carries such a large fraction $z_{\pi}$ of the total energy of the reaction that it does not allow the production of another particle.

  The HERMES and COMPASS collaborations have published $\langle \cos{\, 2 \phi_{\pi}} \rangle$ moments of $\pi^+$ and $\pi^-$ SIDIS, including $z_h$ up to 0.7 \cite{Barone:2009hw}.
  However, it is not possible to make a direct comparison to our $\sigma_{TT}$ $\pi^0$ data as the $\pi^+$ and $\pi^-$ moments on the proton have different signs and magnitudes for Boer-Mulders effect.
  On the other hand, the higher twist Cahn effect, which also contributes to $\sigma_{TT}$, does not give by itself a satisfying description of $\sigma_{TT}$.
  \\
  
  
  \section{Conclusions}
  \label{sec10}
  
  We extracted the separated differential $\pi^0$ cross section at Jefferson Lab, Hall A, at four kinematic settings: Kin2 and Kin3 with a 3\% statistical precision, and Kin$X$2 and Kin$X$3 with a 5\% statistical precision.
  We studied the $Q^2$ dependence of the hadronic tensor with the two first settings, and the $x_{\rm Bj}$ dependence with the latter two.
  
  The shape and order of magnitude of the cross section componants indicate that the $t$-channel meson-exchange model is able to reproduce the total $\pi^0$ cross section, but it would still need improvement for the description of the other components.
  
  Table \ref{XsecHadTens_Q2Wdeps} summarizes the contradiction between our data and the leading twist QCD prediction for high $Q^2$. 
  Instead of an $\sim Q^{-6}$ dependence we find, under the assumption that $\sigma_T$ is negligible (which is very unlikely), a $Q^{-3}$ dependence for $\sigma_L$.
  On the other hand, the cross section extracted may show an analogy with the formalism of SIDIS at the exclusive limit.
  Our $ep \rightarrow ep\pi^0$ data, and the Hall C $ep \rightarrow ep\pi^+$ data are important bases for studying the applicability of the SIDIS concepts to exclusive data.
  To improve the understanding of our data, we have run another $\pi^0$ experiment in Fall 2010, at two beam energies, allowing us to disentangle $\epsilon_L \frac{d\sigma_L}{dt}$ from $\frac{d\sigma_T}{dt}$.
  
  We acknowledge the essential work of the JLab accelerator division and the Hall A technical staff.
  This work was supported by DOE Contract No. DOE-AC05-06OR23177 under which the Jefferson Science Associates, LLC, operates the Thomas Jefferson National Accelerator Facility.
  We acknowledge additional grants from DOE, NSF, and the French CNRS, ANR, and Commissariat {\`a} l' Energie Atomique.
  
 \bibliography{pi0}

  \newpage

  \begin{table*}
    \begin{center}
      \begin{tabular}{|l|c c|c c|}
	\hline
	\hline
	{\small} & \multicolumn{2}{|c|}{\small} & \multicolumn{2}{|c|}{\small} \\
	& \multicolumn{2}{|c|}{{\large $Q^2$ dependence}} & \multicolumn{2}{|c|}{{\large $x_{\rm Bj}$ dependence}} \\
	{\small} & \multicolumn{2}{|c|}{\small} & \multicolumn{2}{|c|}{\small} \\
	\hline
	& Kin3 & Kin2 & Kin$X$3 & Kin$X$2 \\
	\hline
	$N_{\pi^0}$ & 15516 & 23429 & 5952 & 9860 \\
	$N_{\rm gen}$ & $2.14 \times 10^9$ & $2.14 \times 10^9$ &  &  \\
	$\int {\cal L}\,dt$ & $5.10 \times 10^9$ nb$^{-1}$ & $2.99 \times 10^9$ nb$^{-1}$ &  &  \\
	\hline
	$Q^2    $ (${\rm GeV}^2$) & $2.350 \pm 0.002$ & $1.941 \pm 0.010$ & $2.155 \pm 0.268$ & $2.073 \pm 0.001$ \\
	$x_{\rm Bj}                 $ & $0.368 \pm 0.001$ & $0.368 \pm 0.005$ & $0.335 \pm 0.045$ & $0.394 \pm 0.003$ \\
	$W                $ (GeV) & $2.217 \pm 0.004$ & $2.055 \pm 0.012$ & $2.272 \pm 0.072$ & $2.016 \pm 0.008$ \\
	$t_{\rm min}$ (${\rm GeV}^2$) & $-0.173 \pm 0.001$ & $-0.170 \pm 0.005$ & $-0.137 \pm 0.048$ & $-0.199 \pm 0.003$ \\
	$\epsilon               $ & $0.649 \pm 0.002$ & $0.769 \pm 0.003$ & $0.648 \pm 0.001$ & $0.768 \pm 0.003$ \\
	\hline
	$E_0              $ (GeV) & $5.752 \pm 0.001$ & $5.753 \pm 0.001$ & $5.752 \pm 0.001$ & $5.753 \pm 0.001$ \\
	$E^{\prime}       $ (GeV) & $2.348 \pm 0.007$ & $2.937 \pm 0.020$ & $2.321 \pm 0.029$ & $2.951 \pm 0.016$ \\
	$q^{\rm lab}          $ (GeV) & $3.734 \pm 0.007$ & $3.143 \pm 0.017$ & $3.732 \pm 0.009$ & $3.151 \pm 0.014$ \\
	$p_{\pi}^{\rm c.m.}     $ (GeV) & $0.904 \pm 0.002$ & $0.806 \pm 0.007$ & $0.937 \pm 0.043$ & $0.783 \pm 0.005$ \\
	$k_{\gamma}^{\rm c.m.}  $ (GeV) & $0.910 \pm 0.002$ & $0.813 \pm 0.007$ & $0.942 \pm 0.042$ & $0.790 \pm 0.005$ \\
	\hline
	\hline
      \end{tabular}
      \caption{Average quantities weighted with the cross section for the four kinematics of the experiment. Errors are the maximal deviation of the values in the seven $t_{\rm min}-t$ bins, compared to the averages listed.}
      \label{meanvars}
    \end{center}
  \end{table*}
  \begin{table*}
    \begin{center}
      \begin{tabular}{|c c c|c c c|}
	\hline
	\hline
	\multicolumn{3}{|c|}{\small} & \multicolumn{3}{|c|}{\small} \\
	\multicolumn{3}{|c|}{{\large $Q^2$ dependence}} & \multicolumn{3}{|c|}{{\large $x_{\rm Bj}$ dependence}} \\
	\multicolumn{3}{|c|}{\small} & \multicolumn{3}{|c|}{\small} \\
	\hline
	$t_{\rm min}-t$ (${\rm GeV}^2$) & $\sin{\theta_{\pi}^{\rm c.m.}}$ & $\sin^2{\theta_{\pi}^{\rm c.m.}}$ & $t_{\rm min}-t$ (${\rm GeV}^2$) & $\sin{\theta_{\pi}^{\rm c.m.}}$ & $\sin^2{\theta_{\pi}^{\rm c.m.}}$ \\
	\hline
	\multicolumn{3}{|c|}{Kin3} & \multicolumn{3}{|c|}{Kin$X$3} \\
	\hline
	0.0095 & 0.077 & 0.007 & 0.0095 & 0.076 & 0.007 \\
	0.0298 & 0.144 & 0.021 & 0.0297 & 0.143 & 0.020 \\
	0.0546 & 0.194 & 0.038 & 0.0545 & 0.193 & 0.037 \\
	0.0844 & 0.241 & 0.058 & 0.0843 & 0.240 & 0.058 \\
	0.1188 & 0.285 & 0.081 & 0.1188 & 0.284 & 0.081 \\
	0.1583 & 0.328 & 0.108 & 0.1579 & 0.326 & 0.106 \\
	0.2063 & 0.372 & 0.139 & 0.2057 & 0.370 & 0.137 \\
	\hline
	\multicolumn{3}{|c|}{Kin2} & \multicolumn{3}{|c|}{Kin$X$2} \\
	\hline
	0.0094 & 0.085 & 0.008 & 0.0094 & 0.085 & 0.008 \\
	0.0296 & 0.159 & 0.026 & 0.0296 & 0.160 & 0.026 \\
	0.0541 & 0.215 & 0.046 & 0.0542 & 0.216 & 0.047 \\
	0.0839 & 0.267 & 0.071 & 0.0840 & 0.268 & 0.072 \\
	0.1179 & 0.315 & 0.099 & 0.1181 & 0.316 & 0.100 \\
	0.1576 & 0.362 & 0.131 & 0.1579 & 0.364 & 0.133 \\
	0.2050 & 0.410 & 0.168 & 0.2051 & 0.412 & 0.170 \\
	\hline
	\hline
      \end{tabular}
      \caption{Values for $t_{\rm min}-t$, $\sin{\theta_{\pi}^{\rm c.m.}}$ and $\sin^2{\theta_{\pi}^{\rm c.m.}}$, weighted by the cross section.}
      \label{meanvarsbintmint}
    \end{center}
  \end{table*}
  %
  %
  \begin{table*}
    \begin{center}
      \begin{tabular}{|c|c|c|}
	\hline
	\hline
	\multicolumn{3}{|c|}{\small}\\
	\multicolumn{3}{|c|}{{\large $Q^{2}$ dependence}}\\
	\multicolumn{3}{|c|}{\small}\\
	\hline
	& Kin3 & Kin2 \\
	& $x_{\rm Bj}=0.369$, $Q^2=2.350 {\rm GeV}^2$ & $x_{\rm Bj}=0.368$, $Q^2=1.941 {\rm GeV}^2$ \\
	\hline
	$t_{\rm min}-t$ & \multicolumn{2}{|c|}{$d\sigma_T/dt+\epsilon_L d\sigma_L/dt$} \\
	${\rm GeV}^2$ & \multicolumn{2}{|c|}{${\rm nb/GeV}^2$} \\
	\hline
	0.010 & 377 $\pm$ 10 $\pm$ 12 & 571 $\pm$ 10 $\pm$ 24 \\
	0.030 & 381 $\pm$ 12 $\pm$ 12 & 600 $\pm$ 12 $\pm$ 25 \\
	0.054 & 403 $\pm$ 10 $\pm$ 13 & 641 $\pm$ 12 $\pm$ 27 \\
	0.084 & 425 $\pm$ 11 $\pm$ 14 & 673 $\pm$ 15 $\pm$ 28 \\
	0.118 & 418 $\pm$ 11 $\pm$ 14 & 645 $\pm$ 16 $\pm$ 27 \\
	0.158 & 395 $\pm$ 13 $\pm$ 13 & 636 $\pm$ 25 $\pm$ 27 \\
	0.206 & 384 $\pm$ 13 $\pm$ 13 & 628 $\pm$ 36 $\pm$ 26 \\
	\hline
	& \multicolumn{2}{|c|}{$d\sigma_{TL}/dt$} \\
	\hline
	0.010 &  -13 $\pm$ 23 $\pm$ 10 &   17 $\pm$ 19 $\pm$ 13 \\
	0.030 &   38 $\pm$ 26 $\pm$ 24 &  -43 $\pm$ 22 $\pm$ 12 \\
	0.054 &  -25 $\pm$ 22 $\pm$ 11 &  -23 $\pm$ 21 $\pm$ 12 \\
	0.084 &  -26 $\pm$ 25 $\pm$ 13 &  -19 $\pm$ 27 $\pm$ 14 \\
	0.118 &  -75 $\pm$ 24 $\pm$  9 & -103 $\pm$ 30 $\pm$ 21 \\
	0.158 &  -91 $\pm$ 30 $\pm$  8 & -185 $\pm$ 52 $\pm$ 43 \\
	0.206 & -123 $\pm$ 31 $\pm$ 10 & -189 $\pm$ 74 $\pm$ 34 \\
	\hline
	& \multicolumn{2}{|c|}{$d\sigma_{TT}/dt$} \\
	\hline
	0.010 &  -12 $\pm$ 23 $\pm$ 14 &  -39 $\pm$ 19 $\pm$  7 \\
	0.030 &  -25 $\pm$ 27 $\pm$ 15 & -110 $\pm$ 24 $\pm$ 13 \\
	0.054 &  -74 $\pm$ 22 $\pm$  4 & -141 $\pm$ 22 $\pm$ 17 \\
	0.084 &  -64 $\pm$ 25 $\pm$ 14 & -174 $\pm$ 28 $\pm$ 17 \\
	0.118 & -124 $\pm$ 24 $\pm$ 16 & -319 $\pm$ 29 $\pm$ 23 \\
	0.158 & -137 $\pm$ 29 $\pm$ 15 & -352 $\pm$ 45 $\pm$ 53 \\
	0.206 & -134 $\pm$ 30 $\pm$ 15 & -343 $\pm$ 57 $\pm$ 68 \\
	\hline
	& \multicolumn{2}{|c|}{$d\sigma_{TL^{\prime}}/dt$} \\
	\hline
	0.010 &   9 $\pm$ 49 $\pm$ 20 &  31 $\pm$  51 $\pm$ 15 \\
	0.030 & 119 $\pm$ 55 $\pm$ 21 & 136 $\pm$  61 $\pm$ 24 \\
	0.054 & 129 $\pm$ 46 $\pm$ 12 &  61 $\pm$  56 $\pm$ 41 \\
	0.084 & 151 $\pm$ 51 $\pm$ 30 & 123 $\pm$  68 $\pm$ 20 \\
	0.118 & 153 $\pm$ 47 $\pm$ 17 & 120 $\pm$  69 $\pm$ 24 \\
	0.158 &  87 $\pm$ 54 $\pm$ 23 & 142 $\pm$  91 $\pm$ 36 \\
	0.206 & 127 $\pm$ 51 $\pm$ 15 &  76 $\pm$  99 $\pm$ 80 \\
	\hline
	\hline
      \end{tabular}
      \caption{Separated cross-section values from Eq. \eqref{dsigmav_fdsigmadt} (first quoted value) with statistic errors (second quoted value) and systematic errors (third quoted value) for each of the seven considered bins.}
      \label{XsecTab_stat_systErrs_Q2dep}
    \end{center}
  \end{table*}
  %
  %
  \begin{table*}
    \begin{center}
      \begin{tabular}{|c|c|c|}
	\hline
	\hline
	\multicolumn{3}{|c|}{\small}\\
	\multicolumn{3}{|c|}{{\large $x_{\rm Bj}$ dependence}}\\
	\multicolumn{3}{|c|}{\small}\\
	\hline
	& Kin$X$3 & Kin$X$2 \\
	& $x_{\rm Bj}=0.335$, $Q^2=2.155 {\rm GeV}^2$ & $x_{\rm Bj}=0.394$, $Q^2=2.073 {\rm GeV}^2$ \\
	\hline
	$t_{\rm min}-t$ & \multicolumn{2}{|c|}{$d\sigma_T/dt+\epsilon_L d\sigma_L/dt$} \\
	${\rm GeV}^2$ & \multicolumn{2}{|c|}{${\rm nb/GeV}^2$} \\
	\hline
	0.010 & 439 $\pm$ 19 $\pm$ 14 & 635 $\pm$ 17 $\pm$ 26 \\
	0.030 & 437 $\pm$ 22 $\pm$ 14 & 703 $\pm$ 21 $\pm$ 29 \\
	0.054 & 457 $\pm$ 18 $\pm$ 15 & 683 $\pm$ 19 $\pm$ 28 \\
	0.084 & 442 $\pm$ 21 $\pm$ 14 & 688 $\pm$ 23 $\pm$ 29 \\
	0.118 & 466 $\pm$ 22 $\pm$ 15 & 682 $\pm$ 23 $\pm$ 28 \\
	0.158 & 407 $\pm$ 29 $\pm$ 13 & 662 $\pm$ 34 $\pm$ 28 \\
	0.205 & 406 $\pm$ 34 $\pm$ 13 & 591 $\pm$ 44 $\pm$ 25 \\
	\hline
	& \multicolumn{2}{|c|}{$d\sigma_{TL}/dt$} \\
	\hline
	0.010 &   20 $\pm$ 46 $\pm$ 38 &  -26 $\pm$ 30 $\pm$  22 \\
	0.030 &    2 $\pm$ 50 $\pm$ 17 & -100 $\pm$ 37 $\pm$  61 \\
	0.054 &  -28 $\pm$ 43 $\pm$ 15 &  -88 $\pm$ 32 $\pm$  54 \\
	0.084 &  -37 $\pm$ 50 $\pm$ 19 &  -68 $\pm$ 38 $\pm$ 487 \\
	0.118 &  -74 $\pm$ 55 $\pm$ 27 & -170 $\pm$ 40 $\pm$ 562 \\
	0.158 & -188 $\pm$ 80 $\pm$ 27 & -155 $\pm$ 63 $\pm$ 657 \\
	0.205 & -174 $\pm$ 90 $\pm$ 32 & -228 $\pm$ 82 $\pm$ 738 \\
	\hline
	& \multicolumn{2}{|c|}{$d\sigma_{TT}/dt$} \\
	\hline
	0.010 &  -16 $\pm$ 44 $\pm$ 16 &  -63 $\pm$ 33 $\pm$  18 \\
	0.030 &  -44 $\pm$ 50 $\pm$ 32 &  -83 $\pm$ 41 $\pm$  22 \\
	0.054 &  -63 $\pm$ 42 $\pm$ 15 & -153 $\pm$ 36 $\pm$  24 \\
	0.084 & -114 $\pm$ 47 $\pm$  8 & -186 $\pm$ 43 $\pm$  78 \\
	0.118 & -156 $\pm$ 50 $\pm$ 18 & -327 $\pm$ 44 $\pm$ 109 \\
	0.158 & -244 $\pm$ 66 $\pm$ 35 & -247 $\pm$ 65 $\pm$ 141 \\
	0.205 & -124 $\pm$ 69 $\pm$ 42 & -444 $\pm$ 82 $\pm$ 183 \\
	\hline
	& \multicolumn{2}{|c|}{$d\sigma_{TL^{\prime}}/dt$} \\
	\hline
	0.010 &  68 $\pm$  97 $\pm$ 35 & -23 $\pm$  84 $\pm$ 138 \\
	0.030 &  12 $\pm$ 109 $\pm$ 39 & 112 $\pm$ 100 $\pm$ 104 \\
	0.054 & 236 $\pm$  88 $\pm$ 19 &  50 $\pm$  90 $\pm$  63 \\
	0.084 & 126 $\pm$  99 $\pm$ 26 & 211 $\pm$ 104 $\pm$  95 \\
	0.118 & 119 $\pm$  93 $\pm$ 22 &   3 $\pm$ 106 $\pm$ 111 \\
	0.158 & 246 $\pm$ 106 $\pm$ 89 &  78 $\pm$ 136 $\pm$ 126 \\
	0.205 & 177 $\pm$ 104 $\pm$ 30 &  62 $\pm$ 146 $\pm$ 146 \\
	\hline
	\hline
      \end{tabular}
      \caption{Separated cross-section values from Eq. \eqref{dsigmav_fdsigmadt} (first quoted value) with statistic errors (second quoted value) and systematic errors (third quoted value) for each of the first seven bins in $t_{\rm min}-t$ for $1.95 \;{\rm GeV}^2 < Q^2 < 2.25 \;{\rm GeV}^2 $.}
      \label{XsecTab_stat_systErrs_xBdep}
    \end{center}
  \end{table*}
  \begin{table*}
    \begin{center}
      \begin{tabular}{|c|}
	\hline
	\hline
	\\
	{\large $Q^{2}$ dependence}\\
	\\
	\hline
	Kin3 $x_{\rm Bj}$ = 0.368, $Q^2$ = 2.350 (${\rm GeV}^2$) \\
	\hline
	$\frac{W_{xx}-W_{yy}}{2} = [-562 \pm 62 \pm 32] \times \sin^2{\theta_{\pi}^{\rm c.m.}} \cos{2 \phi_{\pi}}$ nb \\
	$\Re e(W_{xz}) = [97 \pm 18 \pm 8] \times \sin{\theta_{\pi}^{\rm c.m.}} \cos{\phi_{\pi}}$ nb \\
	$\Im m(W_{xz}) = [-103 \pm 17 \pm 7] \times \sin{\theta_{\pi}^{\rm c.m.}} \sin{\phi_{\pi}}$ nb \\
	\hline
	Kin2 $x_{\rm Bj}$ = 0.368, $Q^2$ = 1.941 (${\rm GeV}^2$) \\
	\hline
	$\frac{W_{xx}-W_{yy}}{2} = [-1024 \pm 58 \pm 51] \times \sin^2{\theta_{\pi}^{\rm c.m.}} \cos{2 \phi_{\pi}}$ nb \\
	$\Re e(W_{xz}) = [82 \pm 17 \pm 11] \times \sin{\theta_{\pi}^{\rm c.m.}} \cos{\phi_{\pi}}$ nb \\
        $\Im m(W_{xz}) = [-71 \pm 19 \pm 10] \times \sin{\theta_{\pi}^{\rm c.m.}} \sin{\phi_{\pi}}$ nb \\
	\hline
      \end{tabular}
      \caption[]{$\Phi_{\pi}$-dependent hadronic tensor parametrization for constant $x_{\rm Bj}$. The first error is the statistical error, the second is the systematic error.}
      \label{Had_Tens_Q2dep}
    \end{center}
  \end{table*}
  \begin{table*}
    \begin{center}
      \begin{tabular}{|c|}
	\hline
	\hline
	\\
	{\large $x_{\rm Bj}$ dependence}\\
	\\
	\hline
	Kin$X$3 $x_{\rm Bj}$ = 0.335, $Q^2$ = 2.155 (${\rm GeV}^2$) \\
	\hline
	$\frac{W_{xx}-W_{yy}}{2} = [-770 \pm 135 \pm 63] \times \sin^2{\theta_{\pi}^{\rm c.m.}} \cos{2 \phi_{\pi}}$ nb \\
	$\Re e(W_{xz}) = [121 \pm 43 \pm 17] \times \sin{\theta_{\pi}^{\rm c.m.}} \cos{\phi_{\pi}}$ nb \\
	$\Im m(W_{xz}) = [-139 \pm 35 \pm 14] \times \sin{\theta_{\pi}^{\rm c.m.}} \sin{\phi_{\pi}}$ nb \\
	\hline
	Kin$X$2 $x_{\rm Bj}$ = 0.394, $Q^2$ = 2.073 (${\rm GeV}^2$) \\
	\hline
	$\frac{W_{xx}-W_{yy}}{2} = [-1003 \pm 86 \pm 153] \times \sin^2{\theta_{\pi}^{\rm c.m.}} \cos{2 \phi_{\pi}}$ nb \\
	$\Re e(W_{xz}) = [163 \pm 24 \pm 72] \times \sin{\theta_{\pi}^{\rm c.m.}} \cos{\phi_{\pi}}$ nb \\
	$\Im m(W_{xz}) = [-50 \pm 29 \pm 28] \times \sin{\theta_{\pi}^{\rm c.m.} } \sin{\phi_{\pi}}$ nb \\
	\hline
      \end{tabular}
      \caption{$\Phi_{\pi}$-dependent hadronic tensor parametrization for constant $Q^{2}$. The first error is the statistical error, the second is the systematic error.}
      \label{Had_Tens_xBdep}
    \end{center}
  \end{table*}


  \begin{table*}
    \begin{center}
      \begin{tabular}{|c|c|c|c|c|c|c|c|}
	\hline
	\hline
	\multicolumn{8}{|c|}{\small}\\
	\multicolumn{8}{|c|}{{\large $d^4 \sigma / dQ^2 dx_{Bj} dt d\phi_{\pi}$ (pb ${\rm GeV}^{-4}$)}}\\
	\multicolumn{8}{|c|}{\small}\\
	\hline
	$t_{\rm min}-t$ $({\rm GeV}^2)$ & 0.010 & 0.030 & 0.054 & 0.084 & 0.118 & 0.158 & 0.205 \\
	$\phi_{\pi}$ (deg) &  &  &  &  &  &  &  \\
	\hline
	7.5 & 86.15 $\pm$ 7.85 & 89.61 $\pm$ 8.85 & 96.38 $\pm$ 10.50 & 80.59 $\pm$ 14.74 & 28.16 $\pm$ 21.64 & 41.38 $\pm$ 61.01 & -715.48 $\pm$ 314.53 \\
	22.5 & 92.60 $\pm$ 7.63 & 77.79 $\pm$ 8.45 & 80.61 $\pm$ 8.76 & 74.07 $\pm$ 13.41 & 54.94 $\pm$ 13.63 & 62.17 $\pm$ 30.56 & -99.56 $\pm$ 72.94 \\
	37.5 & 98.69 $\pm$ 8.10 & 90.27 $\pm$ 8.51 & 90.50 $\pm$ 8.34 & 93.88 $\pm$ 10.85 & 95.93 $\pm$ 13.41 & 70.65 $\pm$ 20.83 & 98.36 $\pm$ 43.24 \\
	52.5 & 95.41 $\pm$ 7.93 & 103.59 $\pm$ 8.51 & 114.50 $\pm$ 8.21 & 126.75 $\pm$ 10.17 & 117.93 $\pm$ 11.39 & 117.19 $\pm$ 16.01 & 139.95 $\pm$ 23.66 \\
	67.5 & 108.24 $\pm$ 8.28 & 97.84 $\pm$ 8.01 & 114.29 $\pm$ 8.02 & 118.34 $\pm$ 9.42 & 136.07 $\pm$ 10.91 & 116.74 $\pm$ 12.84 & 127.45 $\pm$ 17.11 \\
	82.5 & 98.59 $\pm$ 7.91 & 101.02 $\pm$ 8.34 & 129.87 $\pm$ 8.29 & 129.04 $\pm$ 9.66 & 146.51 $\pm$ 10.23 & 157.22 $\pm$ 13.93 & 147.81 $\pm$ 16.72 \\
	97.5 & 91.64 $\pm$ 7.19 & 130.36 $\pm$ 9.55 & 123.92 $\pm$ 8.25 & 135.90 $\pm$ 9.49 & 160.29 $\pm$ 10.57 & 165.46 $\pm$ 13.51 & 152.67 $\pm$ 14.22 \\
	112.5 & 81.80 $\pm$ 7.21 & 118.51 $\pm$ 8.91 & 120.10 $\pm$ 7.99 & 145.20 $\pm$ 9.45 & 151.23 $\pm$ 10.09 & 157.42 $\pm$ 12.08 & 143.44 $\pm$ 12.22 \\
	127.5 & 104.15 $\pm$ 7.94 & 111.95 $\pm$ 8.49 & 121.61 $\pm$ 7.78 & 122.87 $\pm$ 8.50 & 137.51 $\pm$ 8.58 & 131.98 $\pm$ 9.92 & 128.34 $\pm$ 10.54 \\
	142.5 & 93.66 $\pm$ 7.44 & 127.54 $\pm$ 9.00 & 114.98 $\pm$ 7.25 & 99.98 $\pm$ 7.21 & 104.60 $\pm$ 7.31 & 141.27 $\pm$ 10.78 & 113.38 $\pm$ 10.26 \\
	157.5 & 103.97 $\pm$ 8.03 & 88.87 $\pm$ 7.30 & 99.26 $\pm$ 6.58 & 109.08 $\pm$ 7.64 & 104.67 $\pm$ 7.35 & 98.20 $\pm$ 9.08 & 103.00 $\pm$ 11.04 \\
	172.5 & 90.92 $\pm$ 7.30 & 81.68 $\pm$ 7.15 & 94.38 $\pm$ 6.53 & 100.86 $\pm$ 7.47 & 90.42 $\pm$ 6.79 & 87.99 $\pm$ 8.87 & 92.65 $\pm$ 11.24 \\
	187.5 & 84.70 $\pm$ 6.80 & 100.32 $\pm$ 7.97 & 86.94 $\pm$ 6.30 & 89.97 $\pm$ 6.74 & 80.05 $\pm$ 6.36 & 82.72 $\pm$ 8.52 & 101.53 $\pm$ 11.29 \\
	202.5 & 97.63 $\pm$ 7.66 & 91.05 $\pm$ 7.56 & 90.84 $\pm$ 6.27 & 90.54 $\pm$ 6.71 & 89.67 $\pm$ 6.46 & 110.74 $\pm$ 9.44 & 102.94 $\pm$ 10.49 \\
	217.5 & 79.67 $\pm$ 7.02 & 99.86 $\pm$ 7.57 & 106.18 $\pm$ 7.21 & 116.88 $\pm$ 7.84 & 111.76 $\pm$ 7.34 & 107.67 $\pm$ 8.79 & 133.04 $\pm$ 11.09 \\
	232.5 & 92.30 $\pm$ 7.66 & 105.19 $\pm$ 8.15 & 120.98 $\pm$ 7.46 & 121.08 $\pm$ 8.19 & 121.82 $\pm$ 7.93 & 142.67 $\pm$ 10.09 & 117.56 $\pm$ 9.62 \\
	247.5 & 117.79 $\pm$ 8.60 & 107.01 $\pm$ 8.34 & 106.08 $\pm$ 7.19 & 135.68 $\pm$ 8.99 & 132.76 $\pm$ 8.80 & 149.91 $\pm$ 11.29 & 157.60 $\pm$ 12.06 \\
	262.5 & 84.96 $\pm$ 7.24 & 114.61 $\pm$ 8.69 & 133.27 $\pm$ 8.49 & 126.02 $\pm$ 9.27 & 153.31 $\pm$ 10.42 & 162.73 $\pm$ 12.93 & 181.18 $\pm$ 15.27 \\
	277.5 & 119.19 $\pm$ 8.59 & 118.22 $\pm$ 8.89 & 119.60 $\pm$ 7.89 & 128.61 $\pm$ 9.19 & 141.31 $\pm$ 9.89 & 144.08 $\pm$ 12.87 & 157.05 $\pm$ 15.51 \\
	292.5 & 107.80 $\pm$ 8.17 & 104.11 $\pm$ 8.50 & 119.42 $\pm$ 7.86 & 137.31 $\pm$ 9.89 & 138.58 $\pm$ 10.12 & 130.68 $\pm$ 13.05 & 131.01 $\pm$ 17.41 \\
	307.5 & 100.35 $\pm$ 8.19 & 92.51 $\pm$ 8.16 & 108.65 $\pm$ 7.92 & 126.16 $\pm$ 10.50 & 116.46 $\pm$ 10.96 & 92.60 $\pm$ 14.06 & 119.71 $\pm$ 25.71 \\
	322.5 & 93.52 $\pm$ 7.74 & 97.68 $\pm$ 8.32 & 100.23 $\pm$ 8.36 & 115.51 $\pm$ 11.82 & 90.45 $\pm$ 12.86 & 146.81 $\pm$ 27.39 & 49.27 $\pm$ 34.24 \\
	337.5 & 108.63 $\pm$ 8.83 & 94.20 $\pm$ 9.35 & 106.68 $\pm$ 9.80 & 103.77 $\pm$ 13.15 & 77.24 $\pm$ 17.57 & 49.70 $\pm$ 42.59 & -31.50 $\pm$ 92.90 \\
	352.5 & 87.30 $\pm$ 7.66 & 80.61 $\pm$ 8.34 & 86.20 $\pm$ 9.59 & 105.08 $\pm$ 15.92 & 48.56 $\pm$ 21.05 & 20.91 $\pm$ 45.96 & 0.00 $\pm$ 297.59 \\
	\hline
	\hline
      \end{tabular}
      \caption{Numerical values of experimental cross section $d^4 \sigma / dQ^2 dx_{Bj} dt d\phi_{\pi}$ for each bin in $t_{\rm min}-t$, $\phi_{\pi}$, for $Q^2 = 1.9 \, {\rm GeV}^2$. The errors are statistical errors only. Details on the obtention of those numbers are provided in the text.}
      \label{ExpXsecTab_Kin2}
    \end{center}
  \end{table*}
    
  \begin{table*}
    \begin{center}
      \begin{tabular}{|c|c|c|c|c|c|c|c|}
	\hline
	\hline
	\multicolumn{8}{|c|}{\small}\\
	\multicolumn{8}{|c|}{{\large $d^4 \Sigma / dQ^2 dx_{Bj} dt d\phi_{\pi}$ (pb ${\rm GeV}^{-4}$)}}\\
	\multicolumn{8}{|c|}{\small}\\
	\hline
	$t_{\rm min}-t$ $({\rm GeV}^2)$ & 0.010 & 0.030 & 0.054 & 0.084 & 0.118 & 0.158 & 0.205 \\
	$\phi_{\pi}$ (deg) &  &  &  &  &  &  &  \\
	\hline
	7.5 & 6.22 $\pm$ 9.89 & 6.66 $\pm$ 9.58 & -0.95 $\pm$ 9.54 & 6.96 $\pm$ 14.43 & -21.24 $\pm$ 19.59 & 19.83 $\pm$ 58.48 & 0.00 $\pm$ 83.84 \\
	22.5 & 13.62 $\pm$ 8.98 & -1.88 $\pm$ 9.61 & -0.83 $\pm$ 8.52 & -1.80 $\pm$ 13.36 & -2.54 $\pm$ 11.96 & -28.44 $\pm$ 27.95 & 18.26 $\pm$ 40.13 \\
	37.5 & -4.77 $\pm$ 9.85 & 6.06 $\pm$ 9.55 & 2.85 $\pm$ 8.15 & -8.86 $\pm$ 11.27 & 17.72 $\pm$ 11.97 & 21.20 $\pm$ 20.84 & -7.14 $\pm$ 31.40 \\
	52.5 & -22.18 $\pm$ 10.06 & -2.40 $\pm$ 9.55 & 13.96 $\pm$ 8.39 & -0.88 $\pm$ 10.88 & -6.61 $\pm$ 10.58 & -18.57 $\pm$ 16.35 & 50.01 $\pm$ 17.36 \\
	67.5 & -8.76 $\pm$ 10.42 & 16.95 $\pm$ 8.89 & 7.42 $\pm$ 8.21 & -9.60 $\pm$ 10.25 & 6.93 $\pm$ 10.18 & -13.60 $\pm$ 13.17 & -3.34 $\pm$ 14.12 \\
	82.5 & -5.25 $\pm$ 10.67 & 0.00 $\pm$ 9.03 & -9.81 $\pm$ 8.45 & 3.90 $\pm$ 10.57 & 4.84 $\pm$ 9.71 & 22.12 $\pm$ 14.41 & -37.69 $\pm$ 13.98 \\
	97.5 & 8.59 $\pm$ 9.74 & 16.34 $\pm$ 10.20 & -20.57 $\pm$ 8.36 & 26.79 $\pm$ 10.43 & -0.68 $\pm$ 10.02 & -4.69 $\pm$ 13.89 & 21.67 $\pm$ 11.91 \\
	112.5 & 9.82 $\pm$ 10.23 & -0.72 $\pm$ 9.51 & 5.29 $\pm$ 7.88 & 8.63 $\pm$ 10.73 & -12.19 $\pm$ 9.54 & 18.78 $\pm$ 12.60 & -6.81 $\pm$ 9.93 \\
	127.5 & -2.50 $\pm$ 10.92 & 3.57 $\pm$ 9.16 & 13.88 $\pm$ 7.41 & -6.90 $\pm$ 9.68 & 4.76 $\pm$ 8.16 & -8.10 $\pm$ 10.35 & 13.84 $\pm$ 8.41 \\
	142.5 & -1.66 $\pm$ 10.30 & -13.54 $\pm$ 9.86 & 3.96 $\pm$ 6.62 & 9.76 $\pm$ 8.27 & 10.18 $\pm$ 7.04 & -7.14 $\pm$ 11.30 & 15.87 $\pm$ 8.27 \\
	157.5 & -2.69 $\pm$ 11.50 & 5.39 $\pm$ 8.09 & 2.95 $\pm$ 6.18 & 19.20 $\pm$ 8.79 & -7.88 $\pm$ 7.19 & 6.10 $\pm$ 9.42 & -7.29 $\pm$ 8.50 \\
	172.5 & 12.55 $\pm$ 10.79 & 20.88 $\pm$ 7.92 & -10.16 $\pm$ 5.94 & -16.02 $\pm$ 8.85 & -8.24 $\pm$ 6.77 & 16.64 $\pm$ 9.77 & 10.93 $\pm$ 8.68 \\
	187.5 & 5.75 $\pm$ 9.62 & -0.00 $\pm$ 8.67 & -4.74 $\pm$ 5.87 & 2.99 $\pm$ 7.56 & 4.24 $\pm$ 5.99 & -9.89 $\pm$ 8.62 & 12.12 $\pm$ 9.10 \\
	202.5 & 5.92 $\pm$ 10.56 & -13.81 $\pm$ 8.20 & -4.44 $\pm$ 5.65 & 3.54 $\pm$ 7.61 & 2.37 $\pm$ 6.29 & -28.85 $\pm$ 9.91 & 6.79 $\pm$ 8.04 \\
	217.5 & -10.71 $\pm$ 9.58 & -10.63 $\pm$ 8.47 & 3.24 $\pm$ 6.85 & -16.26 $\pm$ 8.97 & -7.26 $\pm$ 7.02 & -9.08 $\pm$ 9.20 & 3.71 $\pm$ 8.84 \\
	232.5 & 5.40 $\pm$ 10.90 & -15.68 $\pm$ 8.78 & -7.76 $\pm$ 7.15 & 2.58 $\pm$ 9.26 & -3.52 $\pm$ 7.54 & 3.77 $\pm$ 10.55 & -3.27 $\pm$ 8.13 \\
	247.5 & -23.68 $\pm$ 12.10 & -17.05 $\pm$ 8.95 & -5.79 $\pm$ 7.04 & -21.38 $\pm$ 10.05 & -8.87 $\pm$ 8.30 & 11.32 $\pm$ 11.74 & 3.03 $\pm$ 9.85 \\
	262.5 & -7.44 $\pm$ 9.51 & 12.09 $\pm$ 9.22 & 5.55 $\pm$ 8.62 & -0.74 $\pm$ 10.28 & -10.78 $\pm$ 9.85 & -33.46 $\pm$ 13.21 & 7.86 $\pm$ 12.97 \\
	277.5 & -2.49 $\pm$ 11.48 & -5.96 $\pm$ 9.63 & -8.60 $\pm$ 8.08 & -20.65 $\pm$ 10.12 & -9.05 $\pm$ 9.36 & -15.67 $\pm$ 13.30 & -27.89 $\pm$ 13.06 \\
	292.5 & 11.69 $\pm$ 10.39 & -2.32 $\pm$ 9.35 & -9.74 $\pm$ 7.87 & 4.62 $\pm$ 10.75 & -32.10 $\pm$ 9.53 & -20.57 $\pm$ 13.61 & 12.17 $\pm$ 13.95 \\
	307.5 & -10.15 $\pm$ 10.42 & -8.83 $\pm$ 9.20 & 3.56 $\pm$ 8.00 & 4.55 $\pm$ 11.28 & 0.00 $\pm$ 10.17 & -1.99 $\pm$ 14.53 & -5.51 $\pm$ 19.51 \\
	322.5 & -7.31 $\pm$ 9.35 & 8.48 $\pm$ 9.46 & -7.16 $\pm$ 8.36 & 4.66 $\pm$ 12.28 & 4.34 $\pm$ 11.72 & 57.34 $\pm$ 27.10 & -15.68 $\pm$ 21.79 \\
	337.5 & -3.58 $\pm$ 10.91 & -3.86 $\pm$ 10.72 & -7.04 $\pm$ 9.78 & -1.68 $\pm$ 13.42 & -2.41 $\pm$ 15.33 & -93.25 $\pm$ 39.96 & -44.06 $\pm$ 64.96 \\
	352.5 & -9.92 $\pm$ 9.65 & -0.98 $\pm$ 9.64 & -3.94 $\pm$ 9.53 & -6.38 $\pm$ 15.77 & -6.82 $\pm$ 17.74 & -18.77 $\pm$ 41.25 & 95.20 $\pm$ 229.23 \\
	\hline
	\hline
      \end{tabular}
      \caption{Numerical values of experimental helicity dependent cross section $d^4 \Sigma / dQ^2 dx_{Bj} dt d\phi_{\pi}$ for each bin in $t_{\rm min}-t$, $\phi_{\pi}$, for $Q^2 = 1.9 \, {\rm GeV}^2$. The errors are statistical errors only. Details on the obtention of those numbers are provided in the text.}
      \label{ExpXsechdepTab_Kin2}
    \end{center}
  \end{table*}
    
  \begin{table*}
    \begin{center}
      \begin{tabular}{|c|c|c|c|c|c|c|c|}
	\hline
	\hline
	\multicolumn{8}{|c|}{\small}\\
	\multicolumn{8}{|c|}{{\large $d^4 \sigma / dQ^2 dx_{Bj} dt d\phi_{\pi}$ (pb ${\rm GeV}^{-4}$)}}\\
	\multicolumn{8}{|c|}{\small}\\
	\hline
	$t_{\rm min}-t$ $({\rm GeV}^2)$ & 0.010 & 0.030 & 0.054 & 0.084 & 0.118 & 0.158 & 0.205 \\
	$\phi_{\pi}$ (deg) &  &  &  &  &  &  &  \\
	\hline
	7.5 & 42.30 $\pm$ 5.51 & 56.32 $\pm$ 6.40 & 47.89 $\pm$ 5.81 & 53.76 $\pm$ 7.07 & 39.42 $\pm$ 7.95 & 9.15 $\pm$ 11.23 & 36.28 $\pm$ 17.85 \\
	22.5 & 58.86 $\pm$ 6.82 & 51.25 $\pm$ 6.35 & 51.82 $\pm$ 5.84 & 55.51 $\pm$ 6.82 & 32.64 $\pm$ 6.49 & 36.22 $\pm$ 9.29 & 24.49 $\pm$ 13.91 \\
	37.5 & 55.99 $\pm$ 6.54 & 50.20 $\pm$ 6.37 & 53.57 $\pm$ 5.37 & 51.23 $\pm$ 5.81 & 45.17 $\pm$ 5.76 & 58.82 $\pm$ 7.94 & 35.29 $\pm$ 8.23 \\
	52.5 & 51.60 $\pm$ 6.34 & 62.06 $\pm$ 7.00 & 42.86 $\pm$ 4.45 & 49.20 $\pm$ 5.10 & 60.24 $\pm$ 5.42 & 47.96 $\pm$ 5.86 & 46.77 $\pm$ 5.90 \\
	67.5 & 54.11 $\pm$ 6.32 & 59.86 $\pm$ 6.54 & 53.64 $\pm$ 5.20 & 66.28 $\pm$ 5.79 & 57.54 $\pm$ 5.06 & 63.18 $\pm$ 5.80 & 59.44 $\pm$ 6.30 \\
	82.5 & 43.79 $\pm$ 5.47 & 53.80 $\pm$ 6.28 & 63.64 $\pm$ 5.45 & 55.86 $\pm$ 5.19 & 60.93 $\pm$ 5.41 & 63.16 $\pm$ 6.41 & 63.73 $\pm$ 6.79 \\
	97.5 & 55.28 $\pm$ 6.36 & 48.47 $\pm$ 5.78 & 68.73 $\pm$ 5.80 & 60.38 $\pm$ 5.66 & 75.30 $\pm$ 6.15 & 62.20 $\pm$ 6.06 & 66.39 $\pm$ 6.56 \\
	112.5 & 54.48 $\pm$ 6.32 & 58.22 $\pm$ 6.53 & 63.38 $\pm$ 5.67 & 66.92 $\pm$ 5.94 & 74.56 $\pm$ 5.71 & 61.21 $\pm$ 5.72 & 64.68 $\pm$ 5.76 \\
	127.5 & 54.37 $\pm$ 6.22 & 41.12 $\pm$ 5.22 & 61.10 $\pm$ 5.44 & 62.61 $\pm$ 5.62 & 54.38 $\pm$ 4.50 & 59.08 $\pm$ 5.12 & 63.22 $\pm$ 4.92 \\
	142.5 & 60.22 $\pm$ 6.67 & 63.82 $\pm$ 6.48 & 52.46 $\pm$ 4.90 & 59.45 $\pm$ 5.48 & 52.72 $\pm$ 4.53 & 76.04 $\pm$ 5.64 & 55.82 $\pm$ 4.53 \\
	157.5 & 57.59 $\pm$ 6.54 & 54.90 $\pm$ 6.04 & 49.94 $\pm$ 4.87 & 60.05 $\pm$ 5.20 & 59.28 $\pm$ 4.72 & 46.22 $\pm$ 4.44 & 49.73 $\pm$ 4.67 \\
	172.5 & 38.95 $\pm$ 5.26 & 56.20 $\pm$ 6.59 & 53.68 $\pm$ 5.04 & 52.13 $\pm$ 5.21 & 54.33 $\pm$ 4.78 & 49.83 $\pm$ 5.03 & 58.60 $\pm$ 5.40 \\
	187.5 & 56.32 $\pm$ 6.32 & 44.95 $\pm$ 5.85 & 47.40 $\pm$ 4.57 & 52.82 $\pm$ 5.12 & 56.14 $\pm$ 4.94 & 45.60 $\pm$ 4.75 & 51.81 $\pm$ 5.26 \\
	202.5 & 47.94 $\pm$ 5.81 & 32.41 $\pm$ 4.80 & 54.65 $\pm$ 5.22 & 48.15 $\pm$ 4.79 & 56.11 $\pm$ 4.60 & 50.68 $\pm$ 4.69 & 56.62 $\pm$ 4.92 \\
	217.5 & 59.05 $\pm$ 6.66 & 51.08 $\pm$ 6.13 & 57.24 $\pm$ 5.16 & 62.73 $\pm$ 5.63 & 60.22 $\pm$ 4.73 & 57.02 $\pm$ 4.96 & 52.17 $\pm$ 4.35 \\
	232.5 & 48.04 $\pm$ 5.95 & 47.92 $\pm$ 5.60 & 56.26 $\pm$ 5.21 & 59.48 $\pm$ 5.41 & 56.52 $\pm$ 4.70 & 70.51 $\pm$ 5.57 & 66.53 $\pm$ 5.13 \\
	247.5 & 49.20 $\pm$ 5.91 & 48.71 $\pm$ 5.69 & 51.84 $\pm$ 4.91 & 64.47 $\pm$ 5.74 & 63.20 $\pm$ 5.23 & 65.54 $\pm$ 5.57 & 57.32 $\pm$ 5.33 \\
	262.5 & 48.85 $\pm$ 6.10 & 63.63 $\pm$ 6.95 & 54.14 $\pm$ 5.06 & 55.34 $\pm$ 5.34 & 74.01 $\pm$ 5.77 & 63.37 $\pm$ 6.23 & 61.63 $\pm$ 6.12 \\
	277.5 & 53.33 $\pm$ 6.02 & 48.07 $\pm$ 6.01 & 61.57 $\pm$ 5.55 & 70.15 $\pm$ 6.30 & 68.22 $\pm$ 5.50 & 57.00 $\pm$ 5.75 & 59.99 $\pm$ 5.99 \\
	292.5 & 44.06 $\pm$ 5.84 & 55.57 $\pm$ 6.31 & 66.06 $\pm$ 5.66 & 61.62 $\pm$ 5.70 & 60.79 $\pm$ 5.11 & 56.24 $\pm$ 5.59 & 62.01 $\pm$ 6.00 \\
	307.5 & 53.08 $\pm$ 6.20 & 53.08 $\pm$ 5.99 & 62.68 $\pm$ 5.67 & 62.24 $\pm$ 5.71 & 60.01 $\pm$ 5.38 & 61.06 $\pm$ 6.04 & 50.40 $\pm$ 5.76 \\
	322.5 & 54.68 $\pm$ 6.38 & 54.79 $\pm$ 6.35 & 61.38 $\pm$ 5.68 & 60.12 $\pm$ 6.31 & 49.55 $\pm$ 5.68 & 38.16 $\pm$ 6.38 & 50.64 $\pm$ 7.87 \\
	337.5 & 46.01 $\pm$ 5.53 & 43.90 $\pm$ 5.72 & 37.90 $\pm$ 4.73 & 46.75 $\pm$ 6.09 & 45.29 $\pm$ 7.48 & 52.43 $\pm$ 10.29 & 21.25 $\pm$ 9.45 \\
	352.5 & 45.97 $\pm$ 5.84 & 57.70 $\pm$ 6.92 & 46.95 $\pm$ 5.44 & 45.29 $\pm$ 6.71 & 39.98 $\pm$ 7.38 & 30.17 $\pm$ 11.50 & 8.57 $\pm$ 27.98 \\
	\hline
	\hline
      \end{tabular}
      \caption{Numerical values of experimental cross section $d^4 \sigma / dQ^2 dx_{Bj} dt d\phi_{\pi}$ for each bin in $t_{\rm min}-t$, $\phi_{\pi}$, for $Q^2 = 2.3 \, {\rm GeV}^2$. The errors are statistical errors only. Details on the obtention of those numbers are provided in the text.}
      \label{ExpXsecTab_Kin3}
    \end{center}
  \end{table*}
  
  \begin{table*}
    \begin{center}
      \begin{tabular}{|c|c|c|c|c|c|c|c|}
	\hline
	\hline
	\multicolumn{8}{|c|}{\small}\\
	\multicolumn{8}{|c|}{{\large $d^4 \Sigma / dQ^2 dx_{Bj} dt d\phi_{\pi}$ (pb ${\rm GeV}^{-4}$)}}\\
	\multicolumn{8}{|c|}{\small}\\
	\hline
	$t_{\rm min}-t$ $({\rm GeV}^2)$ & 0.010 & 0.030 & 0.054 & 0.084 & 0.118 & 0.158 & 0.205 \\
	$\phi_{\pi}$ (deg) &  &  &  &  &  &  &  \\
	\hline
	7.5 & -4.24 $\pm$ 4.00 & -1.50 $\pm$ 6.31 & 0.00 $\pm$ 6.02 & -9.23 $\pm$ 6.92 & -3.75 $\pm$ 8.07 & -0.00 $\pm$ 10.41 & -0.00 $\pm$ 15.07 \\
	22.5 & 0.00 $\pm$ 5.06 & 0.72 $\pm$ 5.78 & 0.00 $\pm$ 5.94 & -3.26 $\pm$ 6.60 & 5.80 $\pm$ 6.73 & -1.73 $\pm$ 8.00 & 3.65 $\pm$ 12.44 \\
	37.5 & 3.81 $\pm$ 5.34 & 7.21 $\pm$ 5.77 & 13.33 $\pm$ 5.45 & 15.34 $\pm$ 5.73 & 7.19 $\pm$ 5.96 & -12.24 $\pm$ 7.12 & -4.80 $\pm$ 7.83 \\
	52.5 & 5.55 $\pm$ 4.57 & 8.48 $\pm$ 6.37 & 4.87 $\pm$ 4.49 & -0.53 $\pm$ 5.07 & 3.98 $\pm$ 5.69 & 4.27 $\pm$ 5.37 & 9.53 $\pm$ 5.84 \\
	67.5 & 1.18 $\pm$ 4.91 & -0.00 $\pm$ 5.88 & 0.00 $\pm$ 5.22 & 12.24 $\pm$ 5.76 & 11.45 $\pm$ 5.25 & 4.64 $\pm$ 5.43 & 10.59 $\pm$ 6.34 \\
	82.5 & 4.73 $\pm$ 4.43 & -4.11 $\pm$ 5.84 & 0.00 $\pm$ 5.45 & 4.58 $\pm$ 5.20 & 1.01 $\pm$ 5.55 & 7.43 $\pm$ 5.97 & 3.59 $\pm$ 6.88 \\
	97.5 & 2.12 $\pm$ 6.18 & 5.75 $\pm$ 5.33 & 0.51 $\pm$ 5.76 & 13.43 $\pm$ 5.64 & 4.58 $\pm$ 6.27 & 1.17 $\pm$ 5.59 & 15.58 $\pm$ 6.67 \\
	112.5 & 2.92 $\pm$ 6.25 & 5.52 $\pm$ 6.11 & 2.06 $\pm$ 5.63 & -4.29 $\pm$ 5.90 & 11.13 $\pm$ 5.79 & 8.64 $\pm$ 5.22 & 0.00 $\pm$ 5.91 \\
	127.5 & -11.02 $\pm$ 6.67 & 0.00 $\pm$ 4.87 & 9.37 $\pm$ 5.32 & -0.51 $\pm$ 5.56 & 8.43 $\pm$ 4.55 & -0.41 $\pm$ 4.53 & 5.69 $\pm$ 5.06 \\
	142.5 & -6.81 $\pm$ 7.63 & 8.58 $\pm$ 6.36 & 10.28 $\pm$ 4.80 & 11.39 $\pm$ 5.38 & -0.77 $\pm$ 4.53 & 2.28 $\pm$ 4.94 & -0.00 $\pm$ 4.64 \\
	157.5 & -13.29 $\pm$ 7.55 & -2.77 $\pm$ 5.93 & -3.90 $\pm$ 4.80 & 1.43 $\pm$ 5.07 & 4.74 $\pm$ 4.74 & 3.78 $\pm$ 3.81 & -4.98 $\pm$ 4.71 \\
	172.5 & -2.83 $\pm$ 6.75 & 3.61 $\pm$ 6.35 & 1.92 $\pm$ 4.91 & 3.74 $\pm$ 5.12 & -3.29 $\pm$ 4.66 & 2.07 $\pm$ 4.13 & 5.23 $\pm$ 5.44 \\
	187.5 & -5.38 $\pm$ 6.63 & -0.71 $\pm$ 5.67 & 3.54 $\pm$ 4.36 & 3.02 $\pm$ 5.03 & 6.66 $\pm$ 5.00 & -1.61 $\pm$ 3.94 & 3.32 $\pm$ 5.23 \\
	202.5 & -8.93 $\pm$ 5.95 & -4.83 $\pm$ 4.71 & -8.45 $\pm$ 5.13 & -10.90 $\pm$ 4.67 & -6.67 $\pm$ 4.63 & -7.90 $\pm$ 3.93 & -10.26 $\pm$ 5.01 \\
	217.5 & -0.78 $\pm$ 6.85 & 2.08 $\pm$ 5.90 & 4.35 $\pm$ 5.06 & 1.00 $\pm$ 5.48 & 0.00 $\pm$ 4.74 & -11.32 $\pm$ 4.30 & 0.00 $\pm$ 4.44 \\
	232.5 & -4.80 $\pm$ 6.44 & -1.93 $\pm$ 5.34 & 1.43 $\pm$ 5.11 & -4.06 $\pm$ 5.35 & -8.49 $\pm$ 4.74 & 0.42 $\pm$ 5.07 & -7.76 $\pm$ 5.25 \\
	247.5 & 3.58 $\pm$ 5.60 & -5.90 $\pm$ 5.28 & -7.70 $\pm$ 4.87 & 1.07 $\pm$ 5.72 & -7.42 $\pm$ 5.28 & 4.07 $\pm$ 5.15 & -3.10 $\pm$ 5.47 \\
	262.5 & 1.96 $\pm$ 5.30 & 2.11 $\pm$ 6.47 & -5.69 $\pm$ 5.00 & 1.58 $\pm$ 5.33 & -6.25 $\pm$ 5.85 & 1.74 $\pm$ 5.81 & -1.86 $\pm$ 6.28 \\
	277.5 & 1.78 $\pm$ 4.95 & -4.86 $\pm$ 5.47 & -12.96 $\pm$ 5.57 & -10.76 $\pm$ 6.30 & -0.93 $\pm$ 5.64 & -7.87 $\pm$ 5.45 & -5.52 $\pm$ 6.07 \\
	292.5 & 2.65 $\pm$ 4.07 & -0.68 $\pm$ 5.79 & -7.83 $\pm$ 5.70 & -10.24 $\pm$ 5.69 & -7.27 $\pm$ 5.27 & -0.53 $\pm$ 5.25 & 2.88 $\pm$ 6.09 \\
	307.5 & -4.44 $\pm$ 4.73 & -2.59 $\pm$ 5.49 & -5.73 $\pm$ 5.75 & -4.25 $\pm$ 5.65 & 3.41 $\pm$ 5.59 & -7.13 $\pm$ 5.53 & -1.28 $\pm$ 5.70 \\
	322.5 & -1.08 $\pm$ 4.73 & -10.89 $\pm$ 5.68 & 3.20 $\pm$ 5.77 & -11.24 $\pm$ 6.17 & 1.89 $\pm$ 5.85 & 1.63 $\pm$ 5.73 & -8.01 $\pm$ 7.47 \\
	337.5 & -2.52 $\pm$ 4.06 & -10.54 $\pm$ 5.06 & -3.00 $\pm$ 4.80 & 3.05 $\pm$ 5.86 & -4.18 $\pm$ 7.75 & -3.38 $\pm$ 8.94 & -8.23 $\pm$ 8.54 \\
	352.5 & -2.73 $\pm$ 3.47 & -9.09 $\pm$ 5.96 & 5.49 $\pm$ 5.37 & 0.84 $\pm$ 6.38 & 2.62 $\pm$ 7.74 & 0.00 $\pm$ 9.42 & 13.67 $\pm$ 22.30 \\
	\hline
	\hline
      \end{tabular}
      \caption{Numerical values of experimental helicity dependent cross section $d^4 \Sigma / dQ^2 dx_{Bj} dt d\phi_{\pi}$ for each bin in $t_{\rm min}-t$, $\phi_{\pi}$, for $Q^2 = 2.3 \, {\rm GeV}^2$. The errors are statistical errors only. Details on the obtention of those numbers are provided in the text.}
      \label{ExpXsechdepTab_Kin3}
    \end{center}
  \end{table*}
    
\end{document}